\documentclass[aps,twocolumn,10pt]{revtex4-1}

\usepackage{amsmath}
\usepackage{amssymb}
\usepackage{graphicx}% Include figure files
\usepackage{bm}% bold math

\usepackage[utf8]{inputenc}
\usepackage[T1]{fontenc}
\usepackage{mathptmx}
\usepackage{dcolumn}% Align table columns on decimal point
\usepackage{bm}

\usepackage{color}

\usepackage{graphicx} 
\usepackage{float}
\usepackage{float}
\usepackage[caption = false]{subfig}
\usepackage{dirtytalk}

\usepackage[normalem]{ulem} %to strike the words

\usepackage{hyperref}
\usepackage{cleveref}
\usepackage{soul}

\renewcommand{\L}{{\mathcal{L}}}

\begin{document}

\title[A Bird's-Eye View of Naming Game Dynamics]{A Bird's-Eye View of Naming Game Dynamics:\\From Trait Competition to Bayesian Inference}

\affiliation{National Institute of Chemical Physics and Biophysics, R\"avala 10, 10143 Tallinn, Estonia}

\author{Gionni Marchetti}
\email{{\tt gionni.marchetti@kbfi.ee}}

\author{Els Heinsalu} 
\email{{\tt els.heinsalu@kbfi.ee}}
% \altaffio at cs Department, XYZ University.}%Lines break automatically or can be forced with \\

\author{Marco Patriarca}
\email{{\tt marco.patriarca@kbfi.ee}}
%\homepage{\\ {\tt https://hep.kbfi.ee/index.php/Members/MarcoPatriarca}}

\date{\today{}}
%%%%%%%%%%%%%%%%%%%%%%%%%%%%%%%%%%%%%%%%%%%%%%%%%%%%%%%%%%%%%%%

% =============================================================
\begin{abstract}
The present contribution reviews a set of different versions of the basic naming game model, differing in the underlying topology or in the mechanisms regulating the interactions between agents.
We include also a Bayesian naming game model recently introduced, which merges the social dynamics of the basic naming game model with the Bayesian learning framework introduced by Tenenbaum and co-workers.
The latter model goes beyond the fixed nature of names and concepts of standard semiotic dynamics models and the corresponding one-shot learning process, by describing dynamically how agents can generalize a concept from a few examples, according to principles of Bayesian inference.
\end{abstract}
% =============================================================

%%%%%%%%%%%%%%%%%%%%%%%%%%%%%%%%%%%%%%%%%%%%%%%%%%%%%%%%%%%%%%%
%\pacs{81.05.Uw,68.37.-d,73.20-r}

\maketitle

\tableofcontents{}
%%%%%%%%%%%%%%%%%%%%%%%%%%%%%%%%%%%%%%%%%%%%%%%%%%%%%%%%%%%%%%%

%%%%%%%%%%%%%%%%%%%%%%  INTRODUCTION %%%%%%%%%%%%%%%%%%%%%%%%%%%%
\section{Introduction}
\label{introduction}

It has been by now recognized that statistical physics can be used also to investigate various questions relevant for the study of society and culture~\cite{Loreto-2007a,patriarca2020}.
An example is the emergence of social norms or a common cultural background.
%, with no central control, but through reciprocal contacts between people and groups.  
In fact, a hard-science approach to social phenomena, based on the application of dynamical systems theory, statistical mechanics, and complexity theory, represents the rediscovery of a forgotten deep link between social statistics, on one hand, and statistical mechanics, on the other hand~\citep{Ball-2002a,Castellano-2009a}.%, with both quantitative and methodological analogies.

The Naming Game (NG) model and the other models of semiotic dynamics related to it have played a relevant role in the study of cultural diffusion and evolution, since their introduction in the 80's and 90's. 
One of the reasons for this 
is that they offer a simple, yet effective representation of cultural spreading mechanisms.

A main motivation behind the introduction of the NG model was to understand the spontaneously emerging consensus about the use of one or more words in a group of interacting individuals.
This problem is deeply linked to the more general question about the origin of a common language, shared by a group of individuals, and how it can be explained through an underlying ``semiotic dynamics''. 

There are two opposite mechanisms,
shared by many other models of social and cultural dynamics\cite{castellano-2000a},
that characterize the way social interactions take place in the NG models: 
(a) a tendency of interacting individuals to become similar to each other (the so-called ``social influence'');
(b) the presence of noisy elements and random events that diversify individuals from each other.
The outcome of the NG model is either \textit{consensus}, a homogeneous state characterized by a single trait that has prevailed across the whole group of individuals, or \textit{fragmentation}, a heterogeneous state with different groups characterized by different traits.

The NG has evolved along many new directions, becoming a paradigm in various problems.
In fact, it is related to some models of innovation diffusion~\cite{Tuzon-2018a}, language competition (with bilinguals)\cite{Patriarca2012a,patriarca2020}, and opinion dynamics\cite{Castellano-2009a,Sirbu2017}.
From the modeling point of view, the NG can be considered as a cultural competition model between non-excluding options $(A,B,C,\dots)$, describing how different options can spread between interacting individuals when used or be forgotten if unused. 
Like other models the NG model relies on universal mechanisms of cultural spreading, selection, and competition.
The mathematical equations and the computational algorithms of the family of the NG models are often similar or even equivalent to those of other models of social dynamics.
Analogies and differences between different models become more clear considering the mean-field (MF) limit of individual-based models~\cite{Castellano-2009a} (see Sec. \ref{beta_Loreto}).

Based on the type of dynamical rules, models of opinion dynamics and cultural spreading can be categorized in the following way:

(a) Models in which the approach to consensus formation is reached through a direct competition process between different cultural traits that are considered as fixed entities. 
A prototypical example is the family of voter models\cite{castellano-2000a}.

(b) Semiotic dynamics models, which describe how names and concepts link to each other, in a Sussurean sense\cite{Odgen-1923a}, in the minds of the individuals; in these models, different options do not exclude each other and reinforcement processes (and memory effects) can be taken into account.
An example is provided by the Lenaerts model or the NG model. 
It is to be noticed that, in these models, concepts and names are assumed to be fixed entities.

(c) Cognitive models, in which the definition of consensus is understood in a dynamical sense: while names can be assigned in advance, concepts are the outcome of a concept learning process. 
The semiotic dynamics of this class of models goes beyond the Sussurean scheme\cite{Odgen-1923a} and requires a suitable framework for describing quantitatively (1) how and when an agent, who initially does not have any predefined concepts, turns the set of acquired experiences into a concept; and (2) how a group of interacting agents can tune their diversified concepts dynamically in order to reach consensus about e.g. a word, so that all the agents can use and understand the same word. 
An example of such a model is the model of word learning, based on a Bayesian framework, discussed in Sec. \ref{sec:bayesianIntro}.

The goal of the present contribution is to provide a short and self-consistent review of the NG, focusing on how it is employed at different levels of description of socio-cultural processes, from the basic NG model describing direct competition between different traits\cite{castellano-2000a} to the word learning process described within a Bayesian framework of a recently introduced cognitive version of the NG model\cite{Marchetti-2020a}.

This review is limited to a set of selected models and many interesting versions of NG model, proposed over the years, are left out.

%%% PLAN OF THE PAPER %%%
The paper is structured as follows.
In Sec. \ref{history}, we present a concise timeline of the development of some semiotic dynamics models.
In Secs. \ref{basicModel} and \ref{scenarios}, we illustrate different versions of the NG model, starting from the basic version itself and considering extensions in different  embedding topologies -- concise summaries of relevant concepts and terminology of complex networks are given
to maintain the review self-consistent. 
The Bayesian NG model is introduced and discussed in Sec. \ref{sec:bayesianIntro}.
Finally, a short outlook on future research is given in Sec. \ref{conclusions}.

%%%%%%%%%%%%%%%%%%%%%%%%%%%%%%%%%%%%%%%%%%%%%%%%%%%%%%%%%%%%%%%%%%%%%%%%%%%%%%%%

\section{Language as a game: a timeline}
\label{history}

P. H. Matthews, in his \textit{Linguistics, A Very Short Introduction}\cite{Matthews-2003}, writes that ``Human language is, of course, uniquely human". 
This apparently obvious statement actually points out that some relevant properties of language are quite special in nature. 
In fact, language, as we know it in its complexity, seems by now to be indeed a typical trait of the human species only, similarly to other culture-related phenomena such as the systematic development of knowledge and the ability to make technological innovations.
However, other biological species do have some more simple forms of language.
Language is such a peculiar phenomenon, hardly comparable in a straightforward way to any other phenomena, that it has attracted considerable attention of the philosophers and was a subject of study in many places and schools since ancient times, notably the Indian school of linguistics, much before the development of modern linguistics.

From the perspective of language dynamics and modeling, an interesting reference is autobiography of Augustine of Hippo, ``\textit{Confessions}'' , where he suggests a plausible picture of human language and language learning process based on his own personal experience as a child, similar to that of a game for learning words from his elders. 
In the ``\textit{Philosophical Investigations}'', Wittgenstein\cite{Wittgenstein-1953} cites Augustine and elaborates further on the same topic, developing and formalizing various explanatory examples of languages and language learning process.
Wittgenstein referred to these prototypical situations together with the accompanying language learning process as ``Language Games''.
The present review is mostly concerned with the mathematical modeling of language games.

The \emph{Philosophical Investigations} contributed to inspire artificial-intelligence and mathematical models of language.
Luc Steels, inspired by Wittgenstein, implemented the general idea of language as a game\cite{Steels-2004a} by realizing an artificial-intelligence experiment, the \textit{Talking Heads} \cite{Steels-1995a, Steels-1997a, Steels-1998a, Steels-1999b, Steels-1999h, Steels-2011a}.
In the Talking Heads experiment, software-embodied robots can observe some objects in a common environment through digital cameras, with the goal of naming them by inventing words by their own; robots can interact with each other following some pre-assigned interaction rules; a common dictionary emerges eventually, remarkably without direct external control. 

Besides its obvious technological interest, such an experiments is also significant for general linguistics, in that it closely recalls real situations in natural language development.
For example, a relevant fraction of young twins develop autonomous languages to communicate with each other, which are invented by the twins and used only for communications between them, since usually they cannot be understood by others\cite{bakker_1987}.

Steels also introduce a theoretical frameworks referred to as the NG\cite{Steels-1998b,Steels-1998c}, described as an \textit{adaptive} NG, in that the rules regulating the agent's behavior change as the agent's experience grows.
In that model, each agent knows a possibly different lexicon composed of $W$ names and a certain set of $M$ concepts.
Agents can communicate with each other in pairs: one of the two agents in the role of speaker, uttering a name to communicate a certain concept to the other agent, who is in the role of hearer and tries to infer the meaning of the name conveyed.
In this way, agents can learn new names and concepts as well as create or remove links between them.

The idea of a semiotic model of language as a bipartite network of names and concepts, connected to each other through Sussurean-like links, had been studied by Hurford\cite{Hurford-1989a} before Steels' models-- see also the works by Nowak et al.\cite{Nowak-1999a}.
In those models, consensus is achieved through population dynamics and a reproduction advantage for the agents that make more successful communications.
Instead, in the NG of Steels, it is a reinforcement process based on the success of a word (i.e. how many times the hearer inferred the correct meaning of the name conveyed) that makes agents rewind, create, or remove the name$\leftrightarrow$object links. 
The system can converge to consensus, in which the same set of name$\leftrightarrow$object links are used by all agents.
The convergence toward consensus is effectively measured in terms of success rate, given as the fraction of successful communications that have taken place in the system in a given time interval -- in this sense it is a reinforcement model.
The model in its original formulation is still inspiring today, due to its general structure, not fully explored theoretically, yet: in the model, agents can learn new names and see new objects at any moment, move across a spatial topology, and vary in number if the system is open\cite{Steels-1998c}.

A model that closely follows the spirit of Steels' NG model is the model put forward by Lenaerts et al.\cite{Lenaerts-2005a}.
In this model, learning takes place according to the NG rules through mutual interactions between agents, accompanied by a reinforcement process.
The model was studied also in an extended version that describes the evolutionary dynamics of language\cite{Lenaerts-2005a}.

Baronchelli et al.\citep{Baronchelli-2006a,Baronchelli-2006b} introduced what is referred to as the ``basic NG'', a simplified version especially suited for the study of the relaxation process toward a consensus.
This model is discussed in detail in the following sections.
The model is characterized by the existence of many words but only one concept.
In this way, the model looses part of the spectrum of possible problems that can be investigated, e.g. origin of and interaction between synonyms and homonyms, but allows a  study focused on consensus dynamics.
The basic NG represents a good approximation in situations, where many more words than concepts are presents, so that the corresponding semiotic dynamics can be mimicked by parallel single-concept minimal NG processes.

Lipowski and Lipowska have studied some additional versions of the NG model.
They considered a two-agents model designed for a detailed study of the problem of homonymy and synonymy\cite{Lipowski-2009a}; 
models with reinforcement processes and memory effects on adaptive network topologies\cite{Lipowska-2012a}; 
as well as some evolutionary schemes aimed at investigating the Baldwin effect\cite{Lipowski-2008a,Lipowska-2011a} (the influence of linguistic on biological evolution).
Furthermore, Lipowska studied a heterogeneous version of the NG model, in which each speaker's activity depends on the size of the respective vocabulary\cite{Lipowska-2014a}.

%%%%%%%%%%%%%%%%%%%%%%%%%%%%%%%%%%%%%%%%%%%%%%%%%%%%%%%%%%%%%%%%%%%%%%%%%%%%%%%%
%%%  BASIC MODEL
%%%%%%%%%%%%%%%%%%%%%%%%%%%%%%%%%%%%%%%%%%%%%%%%%%%%%%%%%%%%%%%%%%%%%%%%%%%%%%%%

%%%%%%%%%%%%%%%%%%%%%%%%%%%%%%%%%%%%%%%%%%%%%%%%%%%%%%%%%%%%%%%%%%%%%%%%%%%%%%%%

\section{The basic NG model}
\label{basicModel}

In this section we will outline the main features of the basic NG model.
We will focus on some aspects that are relevant for understanding the emergence of consensus in a population of agents.

In particular, it is known that the influence of topology on the NG dynamics is very important. 
For this reason, we present an overview of the dynamics of the basic model on different types of complex networks\footnote{The network architectures presented in the figures or used in the simulations were generated by means of the Python language software package\emph{NetworkX}\cite{Hagberg-2008a}}.

% ===============================================================
\subsection{Complex networks} 
\label{complexNetworks}
 
Complex networks (CNs) are an abstract representations of complex systems, such as Internet, the cell, the World Wide Web, social networks, scientific collaboration networks, or ecological networks\cite{reka2002, Pastor-Satorras-2015}. 
They have complex topologies, regulated by possibly unknown underlying principles, which make them appear randomly structured.
In fact, this was a main reason for developing probabilistic methods for random-graphs\citep{Dorogovtsev-2003}.

A complex network (or graph) is composed by a set of nodes (or vertices) and a set of links (or edges), each link connecting two nodes.
In individual-based models, agents are usually located on the nodes and each link represents some type of interaction between the connected agents.

In the following, we shall consider only undirected networks, i.e.  networks whose pairs of nodes are not ordered, or, equivalently, the networks' links  represent bidirectional interactions between nodes.

Some quantities (network metrics) are particularly useful for characterizing the different underlying topologies of complex networks \cite{reka2002, Pastor-Satorras-2015,Boccaletti-2006a}:
\begin{enumerate}
\item The degree of a node $i$ is is the number $k_i$ of links connecting it to other nodes.
A node that is highly connected, with respect to other nodes, is often called a ``hub''. 
The \emph{degree distribution} $P\left(k\right)$ gives the probability that a node, randomly chosen, has $k$ links; it provides a useful criterion for classifying network topologies, because it has different functional forms in different classes of networks.
The average degree, which provides an estimate of the average connectivity of nodes, is given by $\langle k \rangle = 2n_0/N$, for a network with $N$ nodes and $n_0$ links, and can also be obtained as the first moment of the degree distribution $P\left(k\right)$.
\item Different quantities can measure the tendency of the nodes to cluster (i.e. to be connected to each other), such as e.g. the \emph{global clustering coefficient} (or transitivity), representing the probability that two nodes, connected to another common node, are also connected to each other.
In the following, we mention the local and the average clustering coefficient.
The local (or individual) clustering coefficient \cite{reka2002,barrat-2000b} of node $i$ is\cite{reka2002}
$c_i = 2 E_i/[k_i(k_i-1)] \equiv E_i / E_i^\mathrm{max}$,
where $E_i$ is the total number of links and $E_i^\mathrm{max} \equiv k_i(k_i-1)/2$ the corresponding maximum number of possible connections in the subgraph constituted by the neighborhood of node $i$ (its $k_i$ neighbors). 
The average clustering coefficient $\langle c \rangle$ can be measured by averaging the individual clustering coefficient over the whole network.
\item The concept of shortest path $\ell(i,j)$ between nodes $i$ and $j$ is clearly relevant in the applications of complex networks theory.
The maximum value of $\ell(i,j)$ within the set of the shortest path $\{\ell(i,j)\}$, with $i,j \in (1 \dots N)$, where $N$ is the number of nodes, is termed the \emph{diameter} of the network\cite{Boccaletti-2006a}, in that it provides a measure of its linear size.
Averaging the shortest path over all the pairs of nodes provides the \emph{average shortest path length} 
(or \emph{characteristic path length}),
$\langle \ell \rangle = \sum_{i \ne j} \ell(i,j) / [n_0 (n_0 - 1)]$,
which measures the transport efficiency or overall navigability across a network\cite{Boccaletti-2006a}.
\end{enumerate}

% ========================================================================
\subsection{Basic NG model on fully connected networks}
\label{ngdynamics}

\textit{Fully connected networks}.
The most simple topology of the pair-wise interactions between agents in an individual-based model is that in which each agent can interact with any other agent.
Such a topology is referred to in the literature in various ways, as e.g. that of a group of agents with \emph{all-to-all connections}, located on the nodes of a \emph{fully-connected network} or \emph{complete graph}, or with \emph{homogeneous mixing} \cite{Baronchelli-2006a,Baronchelli-2006d}.
Also, it is the topology on which the MF approximation of a model is usually studied.

\vspace{0.25cm}

The basic (or minimal) NG model, proposed by Baronchelli et al. in 2006 \cite{Baronchelli-2006c, Baronchelli-2008a, Castellano-2009a, Baronchelli-2016a}, was embedded on a fully connected network. 
This stylized agent-based model is deeply rooted in the pioneering \textit{Talking Heads} experiment, performed by Steels \cite{Steels-1995a, Steels-1997a, Steels-1998a, Steels-1999b, Steels-1999h, Steels-2011a}, and in Wittengstein's original idea of the linguistic games \cite{Wittgenstein-1953}. 
Since then, its ability to show how the consensus can spontaneously emerge from the pairwise interactions between the agents has made it a paradigm in the whole field of semiotic dynamics \cite{Castello-2009a} and hence a subject of countless  studies, some of which will be reported in this review.

In this section we describe the basic NG model and its features, which will serve as reference points also in the following sections.

In the basic NG model there are $N$ agents that associate names (or forms) to objects (or concepts).
Through pairwise interactions in a shared environment, agents can learn new associations or select which ones to maintain or discard. 
To this end, every agent owns an inventory -- namely a vocabulary -- that contains the names known to the agent.
Interactions between agents are asymmetrical, in that one agent plays the role of the \emph{speaker}, passing some information to the other agent, who is in the role of \emph{hearer} (see below for details).
This model assumes that there is only one object in the shared environment, implying that all the names stored in the agents' inventories are synonyms.
This excludes the possibility of studying homonymy and its interaction with synonymy, the reason given being that in real-life setting (``words in a context'') homonymy is almost absent \cite{Baronchelli-2016a,komarova-2004a}.
With this limitation, the NG model remains focused on consensus dynamics.

%%% STRATEGY %%%%%%%%%%%%%%%%%%%%%%%%%%%%%
A first element of the NG model is the \emph{scheme} -- also referred to as \emph{strategy} -- used for choosing the agents entering the conversation, i.e. the speaker and the hearer.
In the basic NG, at each time-step, two agents are randomly selected and one of them is randomly chosen as the speaker (who will point to the object and name it), while the other agent will act as hearer (trying to understand the name conveyed by the speaker).
However, other schemes are possible, and provide different results in other models, because the speaker and the hearer selected in a single interaction might experience very different local environments, determined by the network architecture.
In such cases, the strategy for choosing the agents does matter and is a source of asymmetry not only within each interaction but also for the global macroscopic observables.
No such asymmetry is present in the NG dynamics on homogeneous networks, such as fully connected networks and regular lattices, due to their topological homogeneity. 
It is customary to categorizes strategies into three groups\cite{DallAsta-2006b, Baronchelli-2006b, Barrat-2007a}:
\begin{enumerate}
%\begin{itemize}
\item \emph{Direct strategy}: first, the speaker is randomly selected; then the hearer is randomly chosen among the speaker's neighbors.
\item \emph{Inverse strategy}: first, the hearer is randomly selected; then the speaker is randomly chosen among the hearer's neighbors.
\item \emph{Neutral strategy}: a link is chosen with uniform probability among all the existing links; then  with equal probabilities the role of speaker and hearer are assigned to the agents located on the nodes connected by the link selected.
%\end{itemize}
\end{enumerate}
The dependence on the strategy is a typical feature of social dynamics individual-based models: for example, Castellano et al. addressed the (reverse) voter model \cite{krapivsky-1992a, Castellano-2009a} on a generic heterogeneous uncorrelated graph \cite{castellano-2005a}. 
The key observation to understand such a dependence is that if we randomly choose the first agent (node) as speaker, and then we randomly select the second agent (node) who acts as hearer, among the speaker's neighbors, the consequence is that the agents sitting on high-degree nodes will be typically chosen as neighbors. 
In fact, the degree distributions of the speakers and hearers are $P\left(k\right)$ and $kP\left(k\right) /\langle k \rangle$, respectively\cite{DallAsta-2006b}, implying that the large degree nodes will typically act as hearers in a heterogeneous network. 

In the NG model, choosing randomly the first or second agent as speaker produces equivalent results on complete graphs and regular lattices, due to the homogeneity of their topology. 
The details of the scheme used become important when the NG dynamics is studied over complex networks \cite{DallAsta-2006c, Castello-2009a, Baronchelli-2016a}, see Sec. \ref{NGonNetworks}.
%Here, we consider only the direct strategy.
%and its dynamics studied in  the mean-field approximation 
An explicit example of dependence on the strategy is given for the case of scale-free networks in Sec. \ref{NGonNetworks}.

%%% INTERACTION RULES %%%%%%%%%%%%%%%%%%%%%%%%%%%%%%%%%%%%%%%%%%%%%%%%

The next element of the NG model is the set of rules of the interaction between the speaker and the hearer\cite{Baronchelli-2006c}:

\begin{enumerate}

\item   The speaker randomly selects a word from its inventory.
        If the inventory is empty, the speaker invents a new word, which is added to the inventory and selected for the conversation.

\item   The speaker conveys the selected word to the hearer:

        \begin{itemize}

        \item If the hearer's inventory contains the word conveyed, the two agents update their inventories by erasing all the other words, only maintaining the  conveyed word. 
        This process, which represents an agreement -- in practice the two agents \emph{forget} the other words, is termed a communication \emph{success} in the NG.

        \item If the word conveyed is missing from the hearer's inventory, the hearer adds it to the inventory.
        This process, representing in practice a one-shot learning, is termed a communication \emph{failure} in the NG.
    
    \end{itemize}

%\end{itemize}
\end{enumerate}

\begin{figure}
\includegraphics*[scale=0.55]{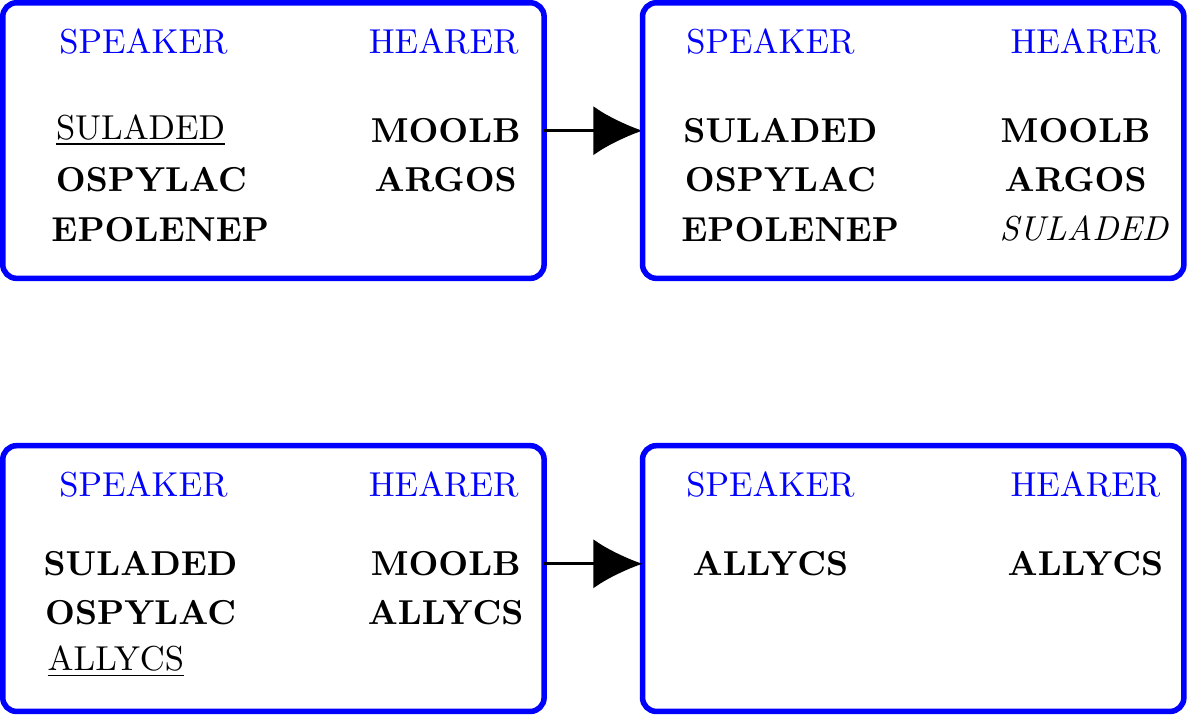}
\caption{Two examples of pairwise interactions according to the basic NG rules. The speaker and hearer's inventories are shown in the cartoon. On the top the speaker randomly selects the underlined word SULADED and conveys it to the hearer. The interaction fails for the hearer's inventory does not contain this name. Thus the hearer must add it to the inventory. On the bottom a successful game is illustrated. The speaker conveys the word ALLYCS that is already present in the hearer's inventory. After this, the agents delete all the words in their inventories and keep only the winning one, i.e. ALLYCS.}
\label{fig:cartoon}
\end{figure}

Pairwise interactions like this one occur at each time step, until the convergence will be achieved, as it is always the case for this basic model.
Top and bottom panels of Fig.~\ref{fig:cartoon} illustrate two possible pairwise interactions which have different outcomes, failure and success, respectively.  

In principle, agents can store \emph{a priori} an unlimited number of words, being the size of agents' inventories not bounded. 
However, it can be proven that within the NG dynamics this never happens and that the system will reach a final (absorbing) state, where all the agents have only one and the same word in their inventories. 

\begin{figure}
\includegraphics*[scale=0.45]{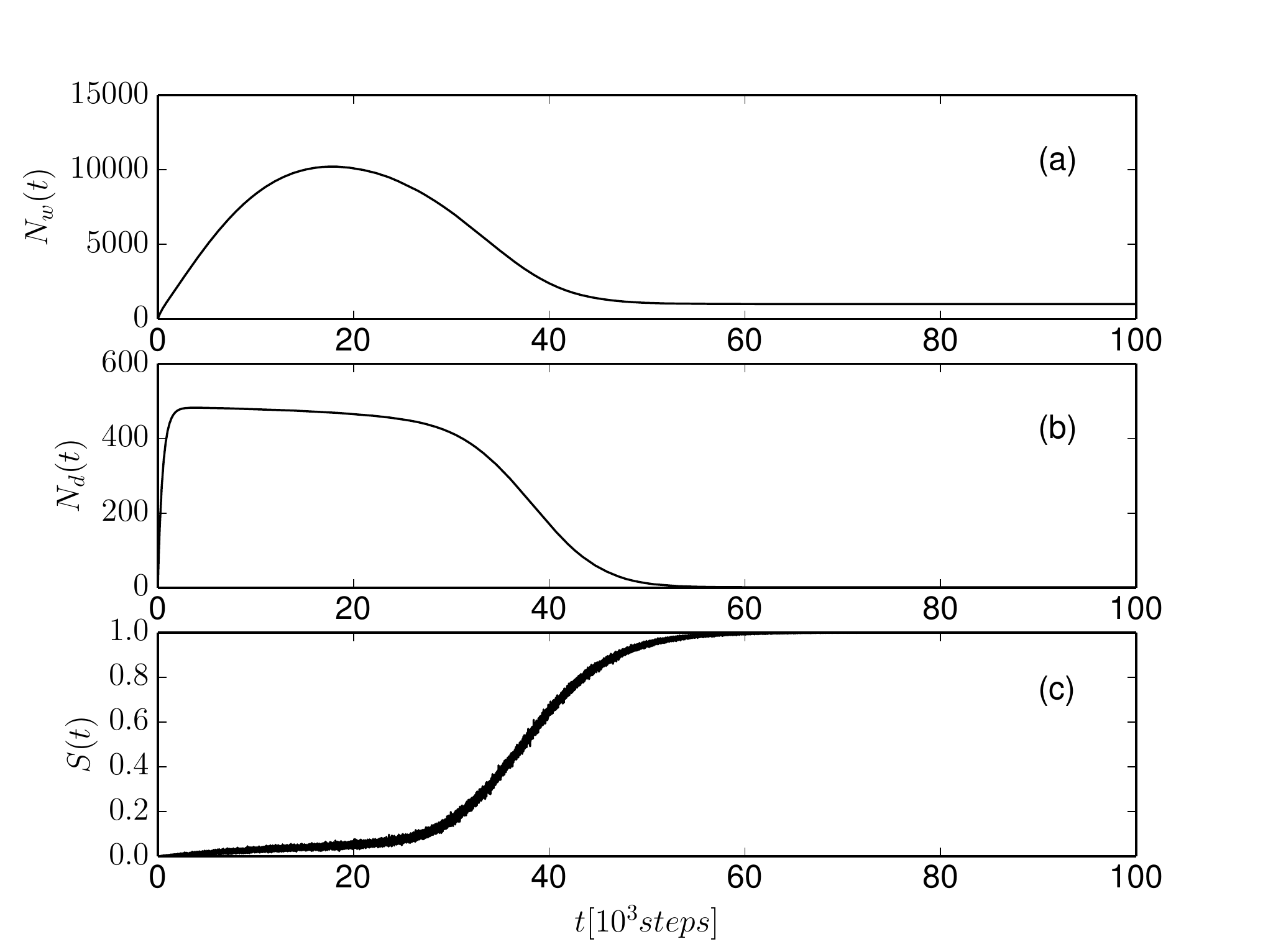}
\caption{The average values of the observables $N_w(t)$, $N_d(t)$, and $S(t)$ as functions of time $t$ obtained from  $1,200$ simulations of the NG dynamics for  a population of $N=1,000$ agents in a fully connected network, see Ref. \cite{Baronchelli-2006c} for comparison. 
Note that  the (average) maximum value of the number of different words is $N_d \approx N/2$.}
\label{fig:fig1}
\end{figure}

In order to understand  the complex processes  leading to the emergence of a final state of global consensus, it is customary to define some time-dependent macroscopic observables that can account for the main dynamical features of the model.
These observables, usually obtained as averages over many different runs of the system, are (a) the success rate $S$, that in each interaction is assigned either the value $S=1$ in case of success or $S=0$ in case of failure; 
(b) the number of different words $N_d$ in the system; 
and (c) the total number of words $N_w$ in the system. 
The latter observable roughly corresponds to the total memory of the system, while the ratio $N_w/N$ is the average amount of memory used by an agent.
The typical time evolution of  $N_w(t)$, $N_d(t)$, and $S(t)$, for a system of size $N=1000$, obtained averaging over $1,200$ realizations, are shown in  Fig.~\ref{fig:fig1}, panels (a), (b), and (c), respectively.
The figure shows that the system reaches a global consensus state, where $N_w(t_\mathrm{conv}) = N$, $N_d(t_\mathrm{conv})=1$, and $S(t_\mathrm{conv})=1$, at the convergence time $t_\mathrm{conv}$. 
This remarkable disorder/order transition occurs spontaneously, without any centralized coordination, showing that in the NG model a population of locally interacting agents is capable of self-organizing, by allowing the formation of a globally shared vocabulary. 
Similar consensus processes are observed in other problems of social sciences, see Ref. \cite{Castellano-2009a}.

The time-evolution of $S$, shown in Fig.~\ref{fig:fig1}, panel (c), gives some important insights about the system dynamics.  
At the beginning, most of interactions are failures, so that the agents play uncorrelated games. 
This implies a linear growth of the observables $N_w$, $N_d$, and $S$. 
Then, there is a second stage, when correlations start to appear, as the agents's inventories have some words in common. 
This gives rise to a collective behavior. 
In the final, third stage, there is a  disorder/order transition that roughly occurs at time $t_\mathrm{max}$, when $N_w$ reaches its maximum, i.e., $ N_w^\mathrm{max} = N_w\left( t_\mathrm{max}\right)$. 
In such a regime, pairwise interactions begin to be successful and $S$ increases monotonically, eventually reaching the unity at consensus.
It can be shown that this transition always takes place, see Ref. \cite{Baronchelli-2006c} for a detailed analysis. 

\begin{figure}
\includegraphics*[scale=0.45]{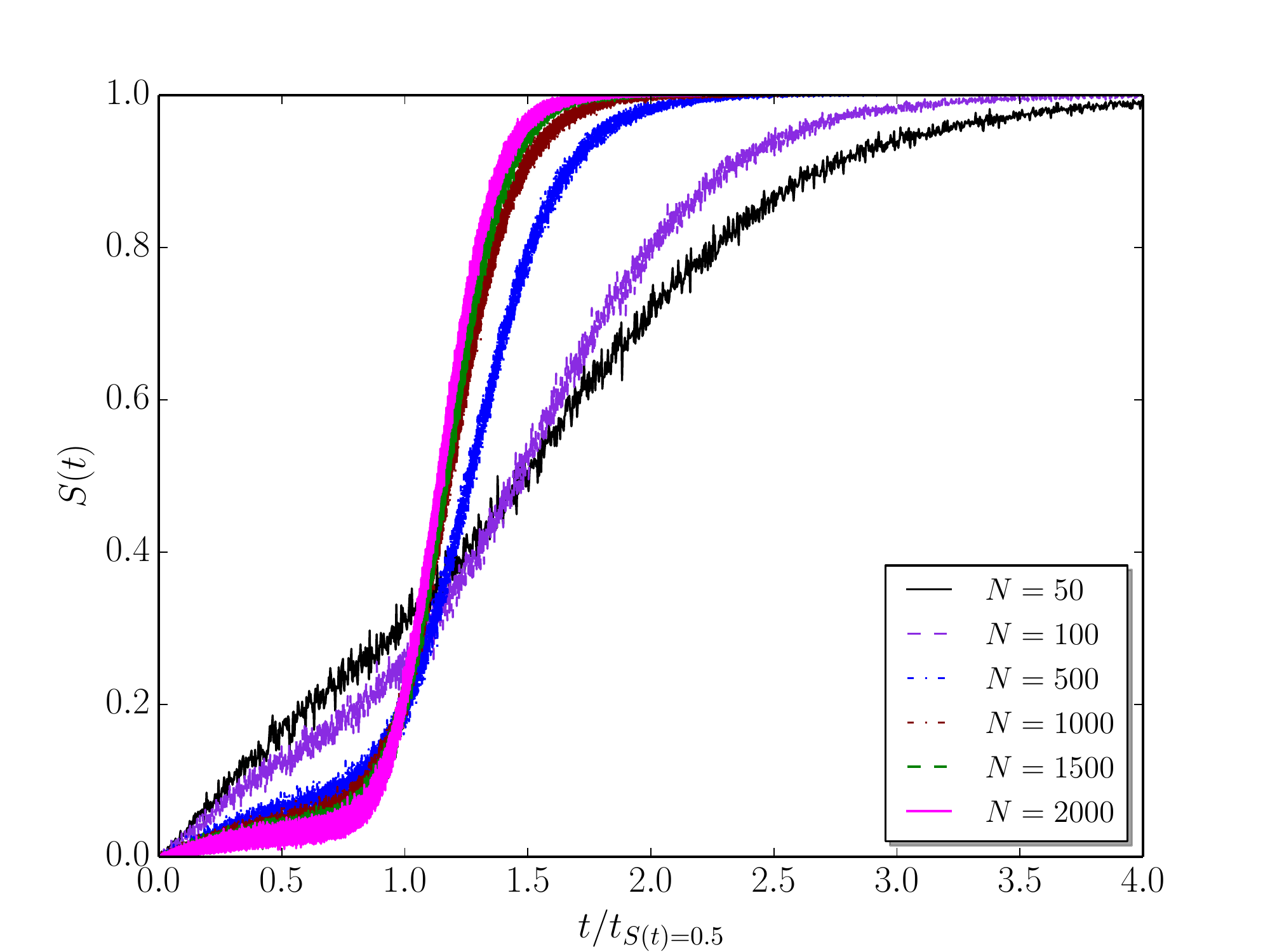}
\caption{Average values of the success rate $S(t)$ as function of the rescaled time $t/t_{S(t) = 0.5}$ for different system sizes $N=50, 100, 500, 1000, 1500, 2000$, obtained from $1,000$ realizations of the NG dynamics on a fully-connected network.}
\label{fig:successRate}
\end{figure}

In Fig.~\ref{fig:successRate}, we plot the success rate $S$ versus time for different system sizes, ranging from $N=50$ to $N=2000$, with time rescaled as $t \to t/t_{S(t) = 0.5} $, the latter being a self-consistent quantity \cite{Baronchelli-2006c}. 
The $S$ curves  clearly show that the disorder/order transition becomes steeper as the number of the agents is increased, hence providing a faster convergence to consensus. 
It is worth noting that the S-shaped curve is also observed in new language conventions spreading in human societies \cite{Baronchelli-2006c, Lass-1997a, Best-2002a, Korner-2002a, Best-2003b}. 
However, the NG dynamics should be understood as a dynamics with time-scales much shorter than those relative to the evolution of a language.

We conclude this section recalling how the macroscopic observables scale with the system size $N$ in the basic NG model. 
It is found that 
$t_\mathrm{conv}$ , $t_\mathrm{max}$, $N_w^\mathrm{max}$, obey  power laws and  scale with $N$ as  
$t_\mathrm{conv} \sim  N^{\alpha} $,  
$t_\mathrm{max}  \sim N^{\beta}$,  
$N_w^\mathrm{max} \sim N^{\delta}$. 
By means of analytical arguments and numerical analysis it can be shown that 
$\alpha \simeq 1.5$, $\beta \simeq 1.5$, $\delta \simeq 1.5$. 
Therefore, the average  amount of memory required by an agent scales as 
$N_w/N \sim N^{\frac{1}{2}}$, which clearly increases with the system size
\cite{Baronchelli-2006c, Baronchelli-2016a}.

% ============================================================================
\subsection{Basic NG model on regular lattices} 
\label{lattices}

\textit{Lattices}.
In regular networks, each node $i$ has the same degree $k_i \equiv \langle k \rangle$ for all $i$.
Regular lattices are particular cases of regular networks: when they are embedded in a real Euclidean space, they form a regular tiling and each node is connected to all its first neighbors.

In the following we consider some particular lattices, namely $1$-dimensional (1D) periodic and $2$-dimensional (2D) square lattices, in which each node has 2 and 4 neighbors, respectively -- in the general case of a (hyper-)cubic lattice of dimension $d$ each node has degree $k = 2d$.
In physics, regular low-dimensional lattices are relevant prototypes of topology for the study of many physical systems with a periodic structures, in which the constituents are usually located at the nodes -- an important example is the Ising model where the constituents are represented by spins\cite{Yeomans-2003}. 

Also in social dynamics, after the fully connected network topology, the natural choice for investigating the influence of different topologies on the dynamics of individual-based models is the topology of low-dimensional regular lattices, e.g. with dimension  $d=1,2,3$.

%%%%%%%%%%%%%%%%%%%%%%%%%%%%%%%%%%%%%%%%%%%%%%%%%%%%%%%%%%
\vspace{0.25cm}

The study of the NG model in a 1D and a 2D lattice was undertaken by Baronchelli et al. \cite{Baronchelli-2006a}. 
Each agents sits on a lattice node and can interacts with $2d$ nearest neighbors only,
%thus establishing multiple interactions with them during the dynamics
a situation that favors local consensus with a high success rate. 

\begin{figure}
\includegraphics*[scale=0.45]{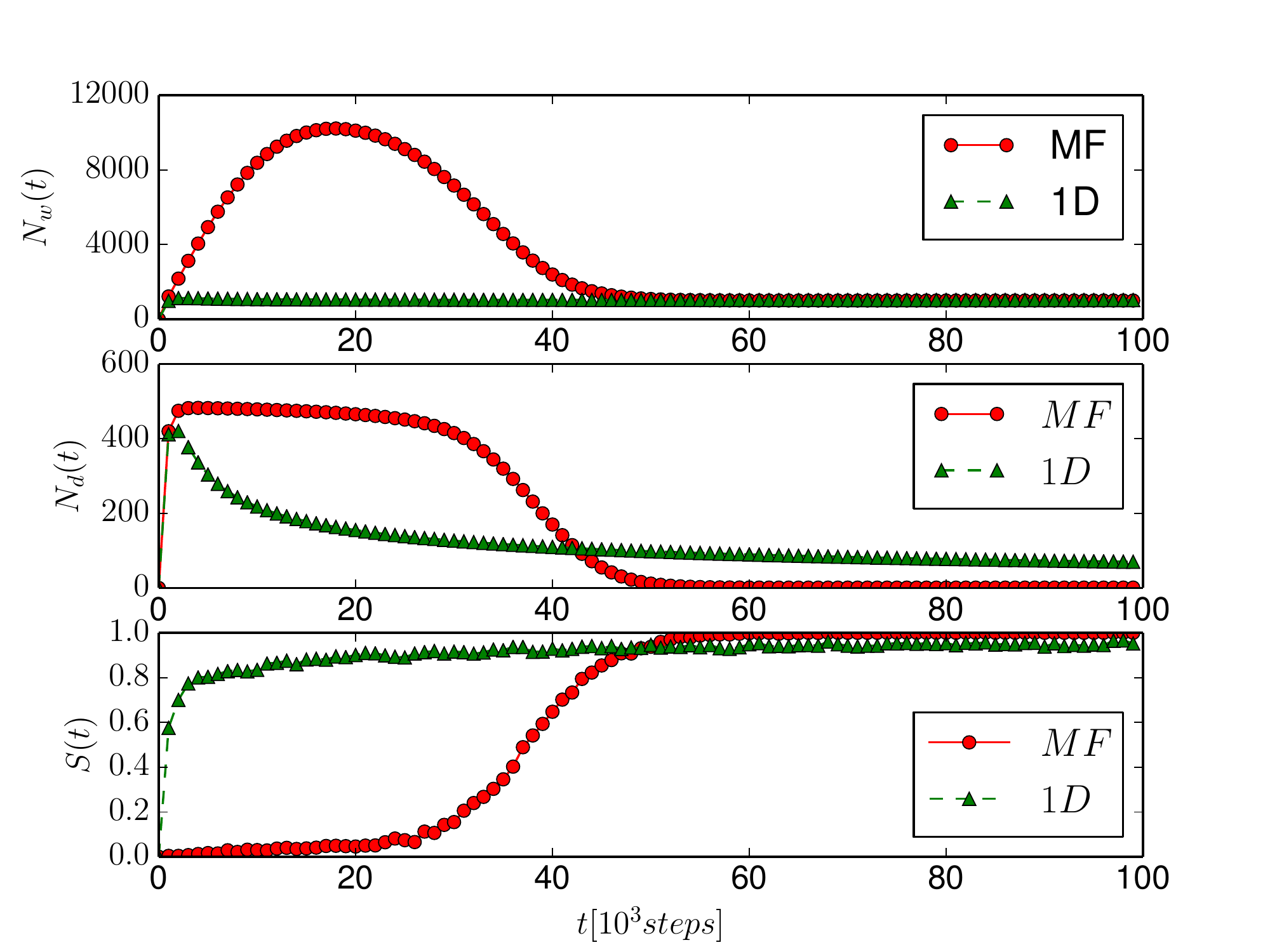}
\caption{Average values of $N_w(t)$, $N_d(t)$, and $S(t)$ as functions of time $t$ obtained from  $1,200$ simulations of the NG dynamics for  a population of $N=1,000$ agents in a fully connected graph (red circles) and on a 1D lattice (green triangles), see Ref. \cite{Baronchelli-2006a} for comparison.}
\label{fig:fig3}
\end{figure}

In Fig.~\ref{fig:fig3}, we plot the time evolution of the main macroscopic observables $N_w$, $N_d$, and $S$, obtained from $1,200$ realizations with $N=1,000$ agents in the 1D case (green triangle symbols) and compare them with the corresponding curves of a fully connected network with the same size (red circle symbols)
--- note that periodic boundary conditions are not assumed in our simulations. 
The differences between the curves clearly show that the convergence to consensus dramatically slows down in the 1D case, but at the same time much less system memory size ($N_w$) is required for consensus;
in particular, one can notice that in the 1D lattice the number of different words $N_d$ lacks the plateau observed in the case of a fully connected network.  
In the 1D case, it is  found that $ t_\mathrm{conv} \sim  N^{3} $, while $ N_{w}^\mathrm{max} \sim  N $. 
The slowing down of the convergence process observed in the 1D case is due to a coarsening phenomenon, namely to the formation of different clusters of neighboring agents who share one word and the consequent competition between the clusters, driven by the fluctuations at their interfaces \cite{Baronchelli-2006a}.

The clusters' interfaces are made by agents with more than one word.
The probability that two neighboring domains are separated by an interface of a given length (measured as the number of its lattice sites) can be studied by constructing a Markov chain. 
In the 1D case, assuming that the interfaces are small point-like objects, it is found that  the probability $\mathcal{P}\left(x,t\right)$ of finding the interface at position $x$ at given time $t$ obeys the following diffusion equation \cite{Baronchelli-2006a},

\begin{equation}\label{eq:diffusion}
\frac{\partial \mathcal{P}(x,t) }{\partial t} 
= 
\frac{D}{N} \frac{\partial^{2} \mathcal{P}(x,t)}{ \partial x^{2}}     \, ,
\end{equation}
where $D\simeq 0.221$, the diffusion coefficient, is in a good agreement with the value $D_{\rm exp} \simeq 0.224$ obtained from numerical simulations. Further analyses of the time-dependence of the domains' size and its relation with the observables $N_w$, $N_d$, and $S$ can be found in Ref. \cite{Lu-2008a}.

Coarsening processes are also observed in the non-equilibrium dynamics of other models, such as the Ising model, and are caused by the dynamics induced by surface tension, referred to also as the curvature-driven dynamics \cite{bray-2002a, DallAsta-2007f, Castellano-2009a}. 
The above picture has been conjectured to hold only in $d$-dimensional lattices with $d \leq 4$, where it can be shown\cite{Baronchelli-2006a, Castello-2009a} that $t_\mathrm{conv}\sim N^{1 + \frac{2}{d}}$. 

In Table~\ref{table:table1}, we report the exponents of the power laws corresponding to $t_\mathrm{conv}$, $t_\mathrm{max}$, and $N_{w}^\mathrm{max}$, which were found for a fully connected network, 1D lattice, and 2D lattice.

% ----------------------------------------------
\begin{table}
\renewcommand{\arraystretch}{2} 
    \caption{Exponents of the power laws for the  NG dynamics in a fully connected network (FCN) and on the 1D and 2D regular lattices \cite{Baronchelli-2006a, Baronchelli-2016a}. 
    Here  $t_\mathrm{conv} \sim N^{\alpha}$, $t_\mathrm{max} \sim N^{\beta}$ and $N_{w}^\mathrm{max}\sim N^{\delta}$. }
\label{table:table1}
%\centering
\begin{ruledtabular}
\begin{tabular}{c c c c   }
Topology & $\alpha$ & $\beta$ &  $\delta$  \\ [1ex]
\hline
    FCN     & $1.5 $         & $1.5 $        & $1.5 $   \\
    1D     & $3.0$          & $1.0$         & $1.0$    \\
    2D     & $2.0$          & $1.0$         & $1.0$     \\ [1ex]
%\hline \hline
\end{tabular}
\end{ruledtabular}
\end{table}
% ----------------------------------------------

% =================================================================================

\subsection{Basic NG model on random graphs and scale-free networks}
\label{NGonNetworks}

A complex underlying topology, such as that of a heterogeneous network, can greatly affect the dynamics of individual-based models embedded in it.
This is true also for the NG dynamics.
In particular, on a complex topology results depend on the choice of the selection strategy adopted (Sec. \ref{basicModel}).
We start by a comparison of the cases of random and scale-free networks, as in the study of NG dynamics on heterogeneous graphs by Dall'Asta et al. \cite{DallAsta-2006b}, which provides useful insights.

%%%%%%%%%%%%%%%%%%%%%%%%%%%%%%%%%%%%%%%%%%%%%%%%%%%%%%%%%%%%%%%%%%%%%%
\vspace{0.25cm}

%%% RANDOM GRAPHS %%%%%%%%%%%%%%%%%%%%%%%%%%%%%%%%%%%%%%%%%%%%%%%%%%%
\textit{Random graphs}.
There are two equivalent definitions of random-graphs (RG)\cite{reka2002}. 
According to Erd\H{o}s  and R\'{e}nyi\cite{erdos-1959a, erdos-1960a, erdos-1961a} (ER model of random network), a random graph is a set of $N$ labeled nodes connected by $n_0$ edges randomly chosen from $N \left(N-1\right)/2$ possible edges. 
Hence, given  $N$  nodes and $n$ edges,  the number of graph realizations is $C_{\left[ N\left(N-1 \right)/2\right]}^{n_0}$ \cite{reka2002}. 
Note that such realizations form a probability space where every graph realization is equiprobable \cite{reka2002}.

\begin{figure}
     \centering
     \subfloat[][]{\includegraphics[width = 1.9in]{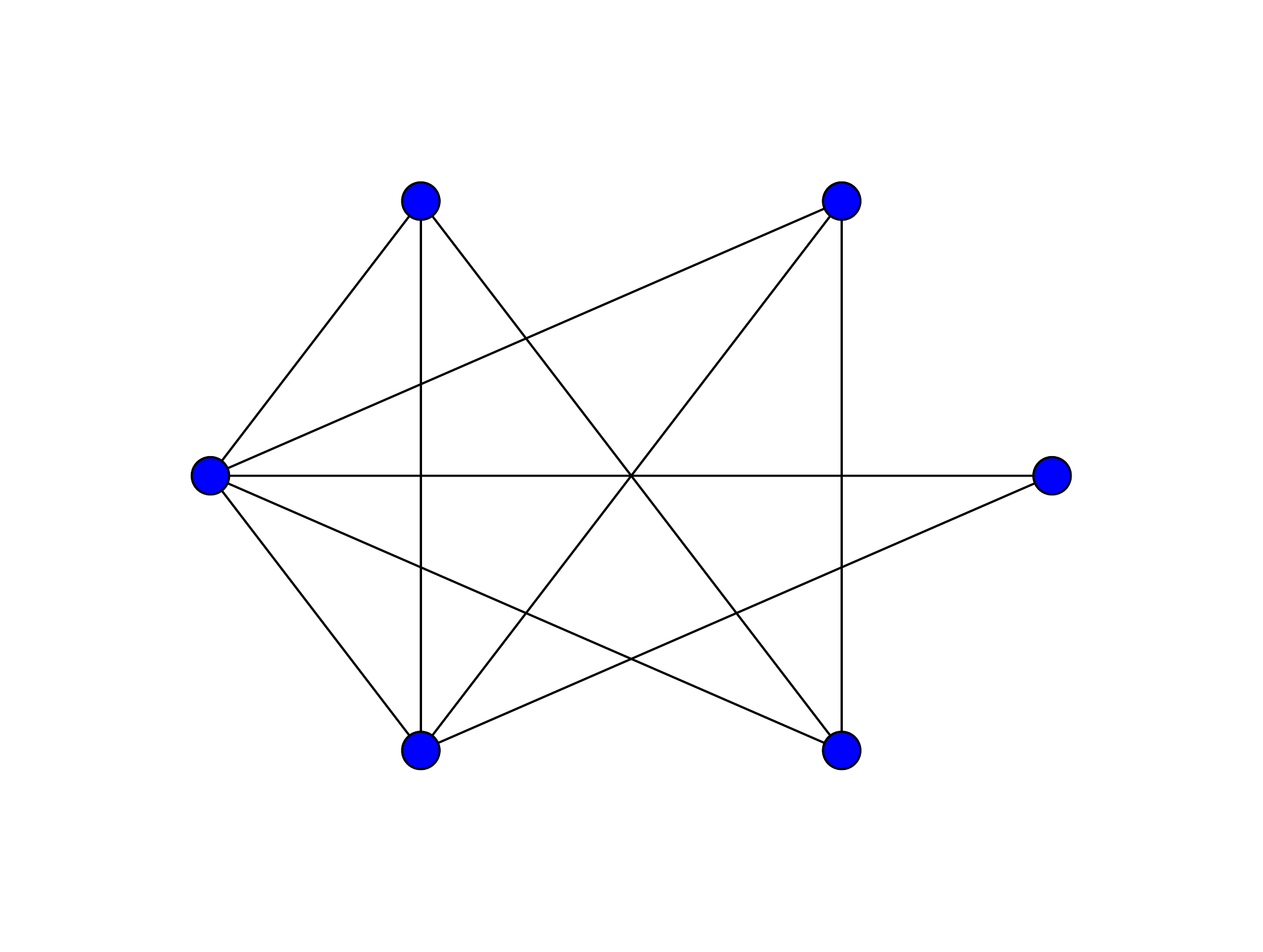}\label{<figure1>}}
     \subfloat[][]{\includegraphics[width = 1.9in]{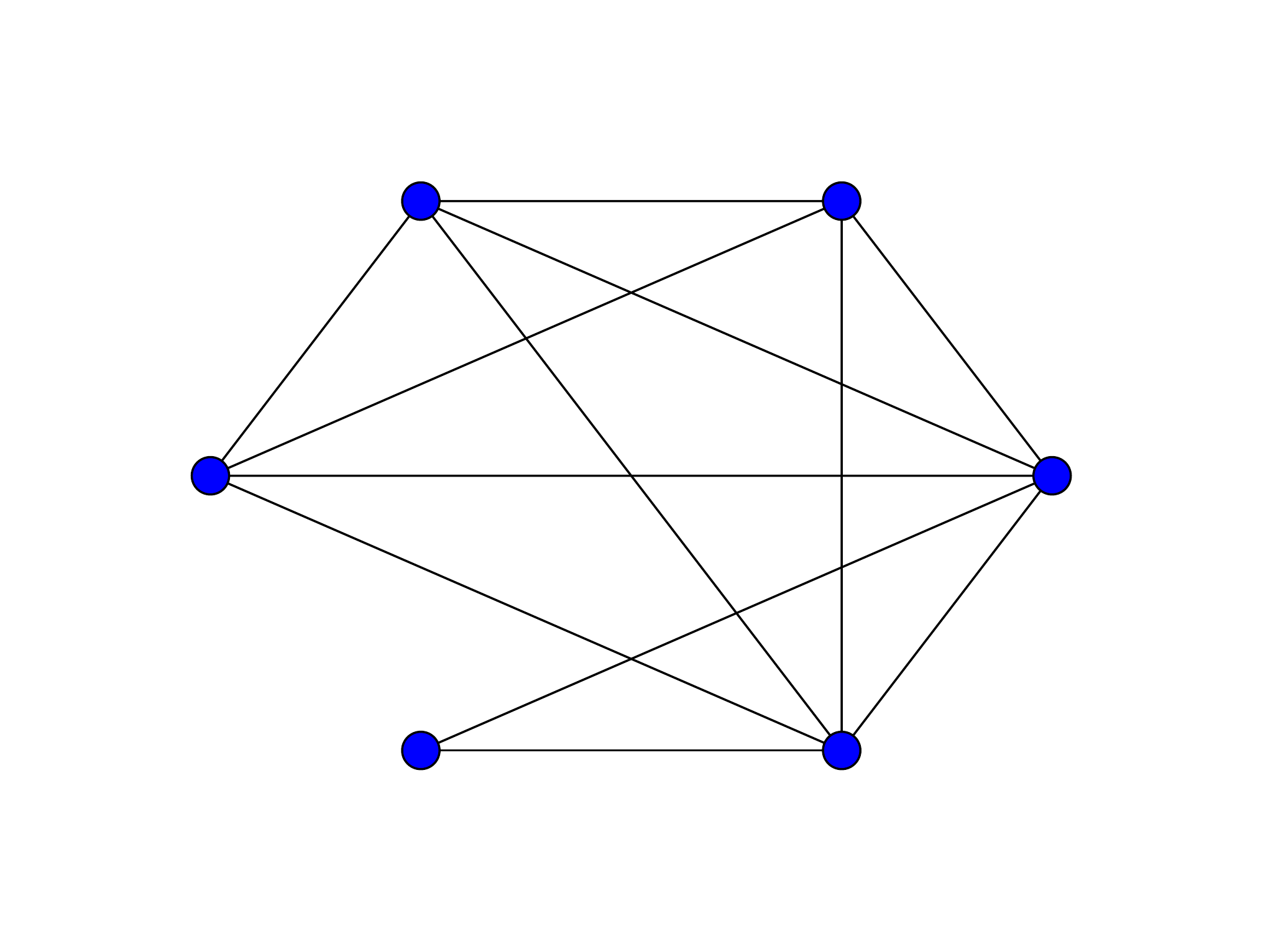}\label{<figure2>}}
\caption{Two realizations of the ER random graph with $N=6$ nodes and connection probability $p_{ER}=0.7$. 
The total number of edges in a random network is $n= p_{ER} N\left(N-1\right)/2$. 
Note that for $p_{ER}=1$ one obtains a fully-connected network.}
\label{fig:ERgraph}     
\end{figure}

The binomial model provides an alternative but equivalent definition of random-graph. 
Starting with $N$ nodes, all pairs of nodes are connected with (uniform) probability $p_{\rm ER}$. 
For example, Fig.~\ref{fig:ERgraph} shows two realizations of ER random graphs:
they are generated using $N=6$ nodes and each pair of nodes is connected with a probability $p_{\rm ER} =0.7$. 
From this model, one expects a graph with a number $n_0 = p_{\rm ER}  N\left(N-1 \right)/2$ of randomly placed links. 
In ER random graphs, nodes have approximately the same number of links close to $\langle k \rangle$ and for a large number $N$ of nodes the degree distribution becomes a Poisson distribution\cite{reka2002},
\begin{equation}\label{eq:poisson}
P\left(k\right) 
    \simeq e^{-p_{\rm ER}N} \frac{\left(p_{\rm ER}N\right)^{k}}{k!} 
    =      e^{-\langle k \rangle} \frac{\langle k \rangle^{k}}{k!} \, ,
\end{equation}
which is characterized by a tail (high-$k$ region) that decreases exponentially. 
Nodes with a connectivity that largely deviates from $\langle k \rangle$ are rare in random graphs. 
In this sense, RGs can be considered as homogeneous networks.

In ER graphs, the mean clustering coefficient is $\langle c \rangle = \langle k \rangle /N=p_{\rm ER}$, which is independent of node's degree, while the mean path length is proportional to the logarithm of network size, $\ell_{\rm rand } \sim \ln N $, a property shared with small-world networks (see Sec.~\ref{NGonSW}).

%%%%%%%%%%%%%%%%%%%%%%%%%%%%%%%%%%%%%%%%%%%%%%%%%%%%%%%%%%%%%%%%%%%%%%
\vspace{0.25cm}

%%%%  SCALE-FREE NETWORKS %%%%%%%%%%%%%%%%%%%%%%%%%%%%%%%%%%%%%%%%%%%%
\textit{Scale-free networks}.
The ER model and the Watts and Strogatz model (see Sec. \ref{NGonSW}) of network cannot describe some topological properties observed in many real technological, biological and social networks.
Indeed, it is found\cite{Barabasi-1999a, reka2002, barabasi-2004} that in many cases networks are characterized by a power law distribution, i.e., $P\left(k\right) \sim k^{-\gamma}$ at large values of $k$, where $\gamma$ is the degree exponent, whose values is usually between 2 and 3.
Due to this special dependence on $k$, these graphs are called \emph{scale-free} networks.
This shape implies that only a few nodes (hubs) have a large number of links, while the majority of nodes have only a few links. 
 
Barab\'{a}si and Albert proposed a model of scale-free networks, considering a dynamical evolutionary origin, jointly caused by two process: growth and preferential attachment. 
Indeed,  it is observed that many real networks grow by continuous addition of new nodes. 
Moreover, the new nodes have higher probability to be connected to those with large number of links. 
The latter process is called preferential attachment, see the algorithm illustrated below. 
On the contrary, no such process is present in the procedures for generating random-graphs or small-world networks (Sec. \ref{NGonSW}), where the connecting probability between nodes is independent of nodes' degree.

\begin{figure}
%\begin{center}
\includegraphics*[scale=0.4]{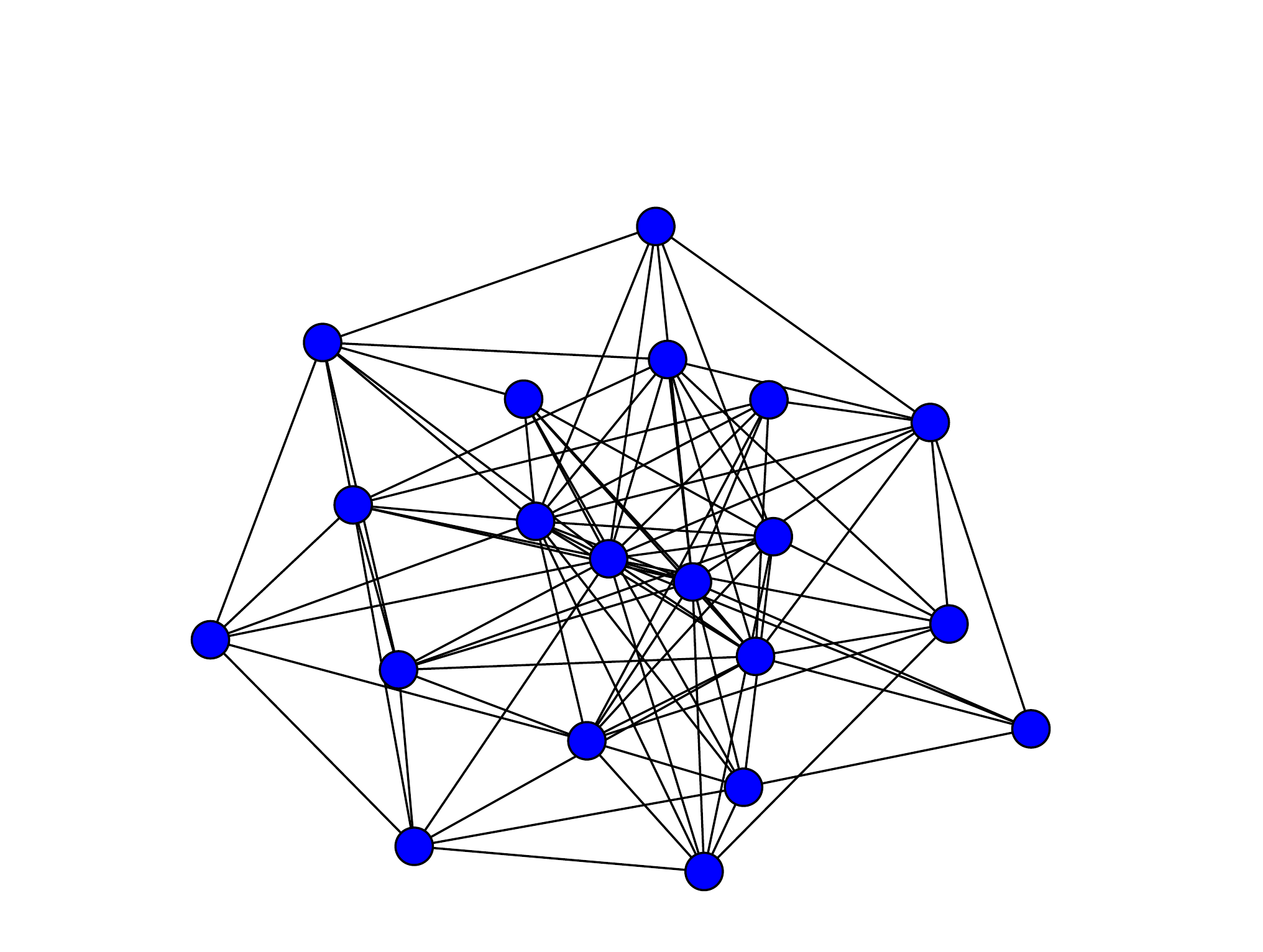}
%\end{center}
\caption{Scale-free network with $N=20$ nodes constructed using the Barab\'{a}si and Albert preferential attachment algorithm with parameter $m = 7$.}
\label{fig:sfNetwork}
\end{figure}

The Barab\'{a}si and Albert (BA) model generates a scale-free networks by means of the following preferential attachment algorithm: \cite{reka2002}

\begin{enumerate}
    \item Initially there is a small number $m_0$ of nodes connected to each other through $m$ edges ($m \leq m_0$).
    \item At each new time-step, a new node $j$ is added.
    \item The new node is then connected to another existing node $i$. The choice of the node $i$ is done with a preferential attachment probability $\Pi$ proportional to the respective connectivity $k_i$,
            \begin{equation}\label{eq:preferential}
                \Pi\left(k_i \right) = \frac{k_i}{\sum_j k_j} \, .
            \end{equation}
    \item Time is increased of one step and the procedure is restarted from point 2 above.
\end{enumerate}

A scale-free network with $N= 20$ nodes, generated by means of the BA preferential attachment algorithm with parameter $m=7$, is shown in Fig.~ \ref{fig:sfNetwork}.
Note the cliques spontaneously emerging, despite the small size of the network.  

The BA model allows the construction of the scale-free networks characterized by power law distributions with $\gamma =3$, independent of the value of the parameter $m$.\cite{reka2002}
Additionally, it is found\cite{reka2002} that the average path length  $\langle \ell \rangle$ is shorter than in the corresponding value $\ell_{\rm rand }$ for random networks, for any $N$, while the average clustering coefficient follows a power law $\langle c \rangle = N^{-0.75}$. 
Clearly, the scale-free networks are heterogeneous, for the low-degree nodes are far more abundant than those with a high degree. 

It is worth noting that a recent study by Broido and Clauset claims that the scale-free architecture is rare among the real-world networks, undermining the idea that the scale-free nature is an underlying principle of the most complex networks \cite{Broido-2019a}. 
By means of a data-driven approach, the authors have shown that out of nearly $1,000$ network data sets, only $4\%$ of them exhibits the strongest-possible evidence of a scale-free architecture. 
However, these statistical tests are applied to finite-size real networks while the scale-free character of the BA model is rigorously valid in the limit of infinite-size networks, see Ref. \cite{Holme-2019a} for an interesting discussion about this apparent issue.

%%%%%%%%%%%%%%%%%%%%%%%%%%%%%%%%%%%%%%%%%%%%%%%%%%%%%%%%%%%%%%%%%%%%%%
\vspace{0.25cm}

%%%%  NG ON RANDOM GRAPHS & SCALE-FREE NETWORKS 
Dall'Asta et al. performed simulations of the NG dynamics on different realizations of ER random graphs and BA scale-free networks. 
For the latter network model, results were also compared with those of uncorrelated scale-free networks constructed by means of the uncorrelated configuration model, but no significant differences were found. 

\begin{figure}
%\begin{center}
\includegraphics*[scale=0.4]{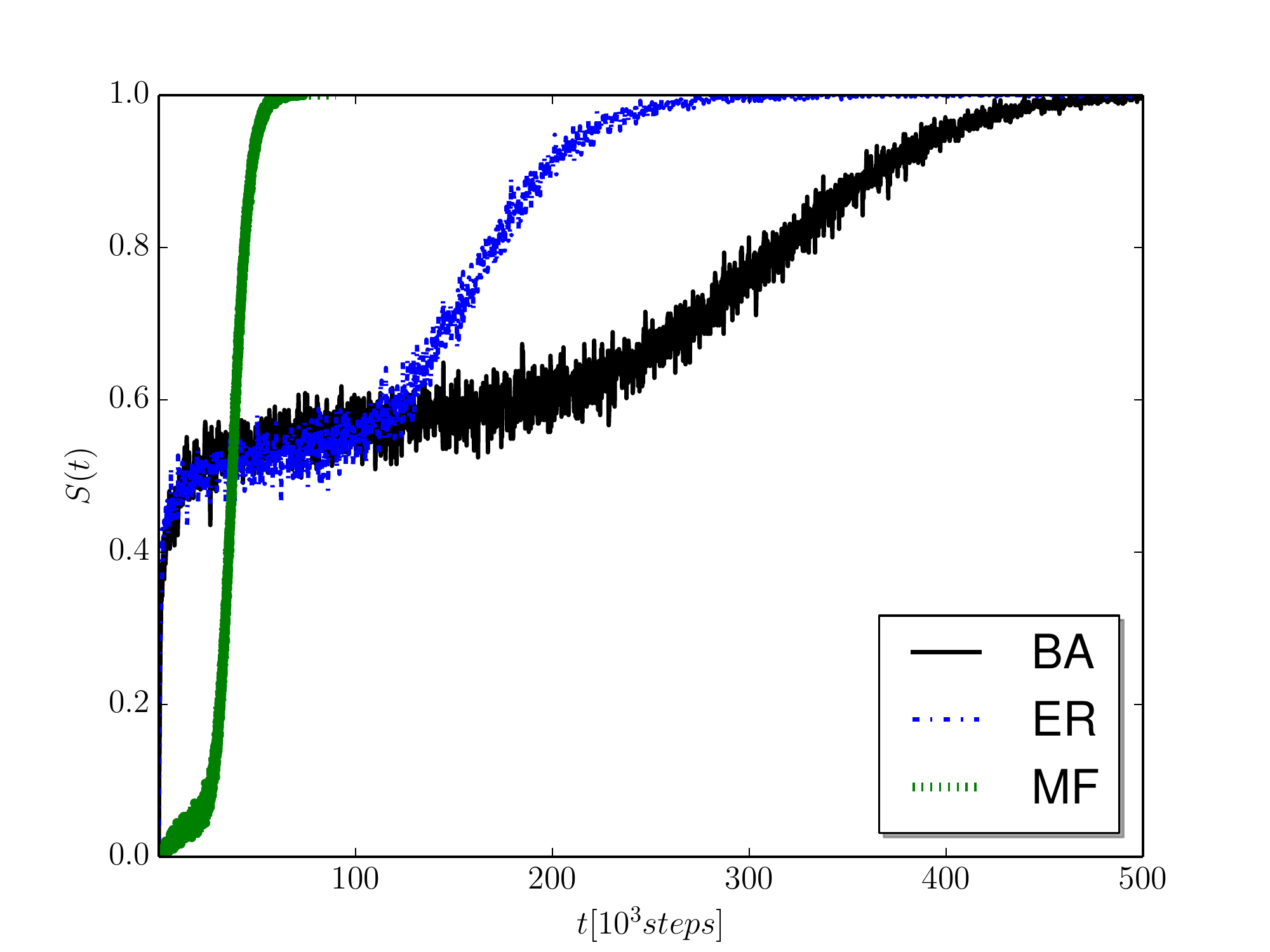}
%\end{center}
\caption{Comparison of the success rate $S(t)$ as a function of time $t$ on a Barab\'{a}si and Albert free-scale network (``BA'', black solid line) and an Erd\H{o}s-R\'{e}nyi random graph (``ER'', blue dashed line).
In both cases the networks have an average degree $\langle k \rangle = 4$, the population is $N=1,000$ agents, and results were averaged over $600$ realizations.
For comparison, also $S(t)$ for a complete graph is shown (``MF'', green dotted line, averaged over $1,200$ realizations).}
\label{fig:successER_BA}
\end{figure}

In Fig.~\ref{fig:successER_BA} the success rate $S(t)$ versus time $t$ for ER (dashed line) and BA (solid line) networks is shown.
Both networks have an average degree $\langle k \rangle = 4$ and a population of $N=1,000$ agents. 
For comparison, the corresponding curve for a complete graph is plotted. 
The success rate curves for the dynamics on complex networks show similar behaviors characterized by an initial linear behavior and a plateau at intermediate times, which is not observed in the case of a complete graph. 
The faster initial growth of $S(t)$, with respect to the case of the complete graph, is due to the finite average degree\cite{DallAsta-2006b}, which is $\langle k \rangle = 4$.

\begin{figure}
%\begin{center}
\includegraphics*[scale=0.4]{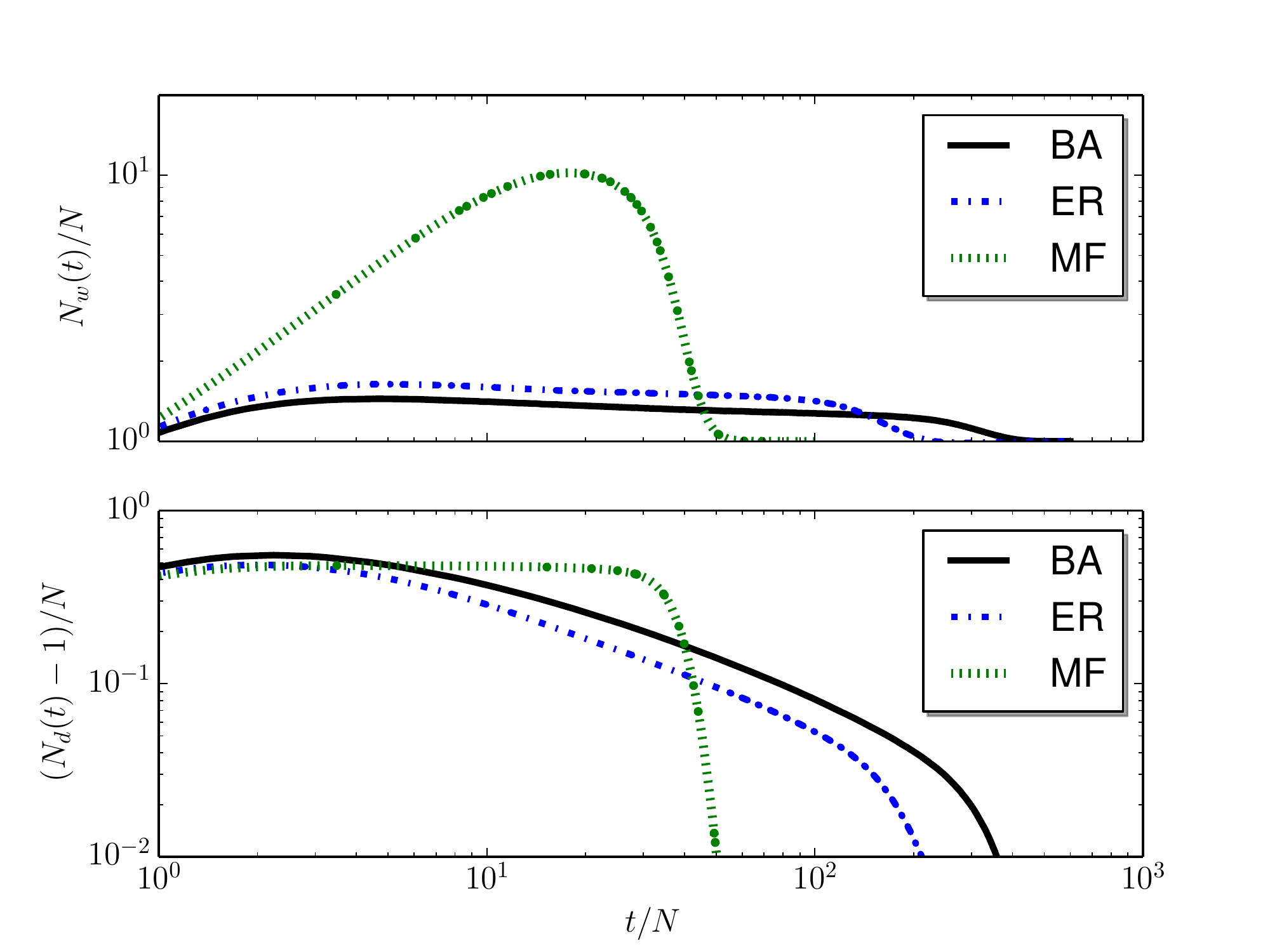}
%\end{center}
\caption{Comparison of the global observables $N_w(t)/N$ (top panel) and $(N_d(t)-1)/N$ (bottom panel) as functions of the rescaled time $t/N$ of a BA network (black solid line), an ER network (blue dashed line), and a complete graph (green dotted line).
The population size is $N=1,000$ agent.
For both the BA and ER networks the average degree is $\langle k \rangle = 4$ and results were averaged over $600$ realizations; results of the complete graph were obtained from $1,200$ realizations.}
\label{fig:ER_BAdynamics}
\end{figure}

Figure~\ref{fig:ER_BAdynamics} compares the average memory per agent in the system, given by the macroscopic observables $N_w/N$ (top panel), and the number of different words per agent, given by the quantity $(N_d-1)/N$ (bottom panel), as functions of the rescaled time $t/N$, for the ER network, the BA network (both networks have average degree $\langle k \rangle = 4$), and the complete graph. 
The memory $N_w/N$ obtained from the NG dynamics on the complex networks, after a sudden increase, reaches a plateau that is lacking in the curve for the complete graph, which instead presents a peak (top panel). 
Notice that the plateaus for the complex networks do not correspond to steady states.
Instead, they represent a dynamical regime in which the system is eliminating more and more names, according to the NG agreement rule.
This becomes evident by looking at the number of different words $(N_d-1)/N$ (bottom panel). 
Moreover, it is  found\cite{DallAsta-2006b} that the length of the plateau increases with the system size $N$.

An extensive analysis of the NG dynamics on ER and BA networks shows that the convergence time $t_\mathrm{conv}$ scales with $N$ as $t_\mathrm{conv} \sim N^{\alpha}$ with $\alpha = 1.4 \pm 0.1$.
This represents a convergence faster than that found for the case of the regular lattices, see Table~\ref{table:table2} for the list of the values of the exponents obtained from the simulations on the various model network architectures and Table~\ref{table:table1} for comparison.
For the sake of completeness, also the exponents $\delta$ for the time of the system memory peak, $N_{w}^\mathrm{max} \sim N^{\delta}$, are listed.
%we report the numerical results for the scaling laws relative to the ER and BA complex networks. 
Dall'Asta et al. showed that this scaling represents a robust feature unaffected by the clustering, average degree $\langle k \rangle$, or degree distribution\cite{DallAsta-2006b, Loreto_2011a} -- see also Ref. \cite{Barrat-2007a} for further details.
In the case of the BA networks, this behavior was confirmed by comparison with the uncorrelated scale-free networks, created according to the uncorrelated configuration model.

\begin{table}
\renewcommand{\arraystretch}{2} 
\caption{The scaling laws for convergence time $t_\mathrm{conv} \sim N^{\alpha}$ and maximum memory $N_{w}^\mathrm{max}\sim N^{\delta}$ for the  basic NG on the ER graphs, BA and WS network models.}
\label{table:table2}
%\centering
\begin{ruledtabular}
\begin{tabular}{c c c c   }
%\hline \hline
Network Model & $\alpha$ & $\delta$ &  Reference  \\ [1ex]
\hline
    ER         &$1.4 \pm 0.1$        & $1.0$         & Ref. \cite{DallAsta-2006b} \\
    BA         &$1.4 \pm 0.1$        & $1.0$         & Ref. \cite{DallAsta-2006b} \\ 
    WS         &$1.4 \pm 0.1$        & $1.0$        & Ref. \cite{DallAsta-2006a} \\ [1ex]

%\hline \hline
\end{tabular}
\end{ruledtabular}
\end{table}

\begin{figure}
\includegraphics*[scale=0.4]{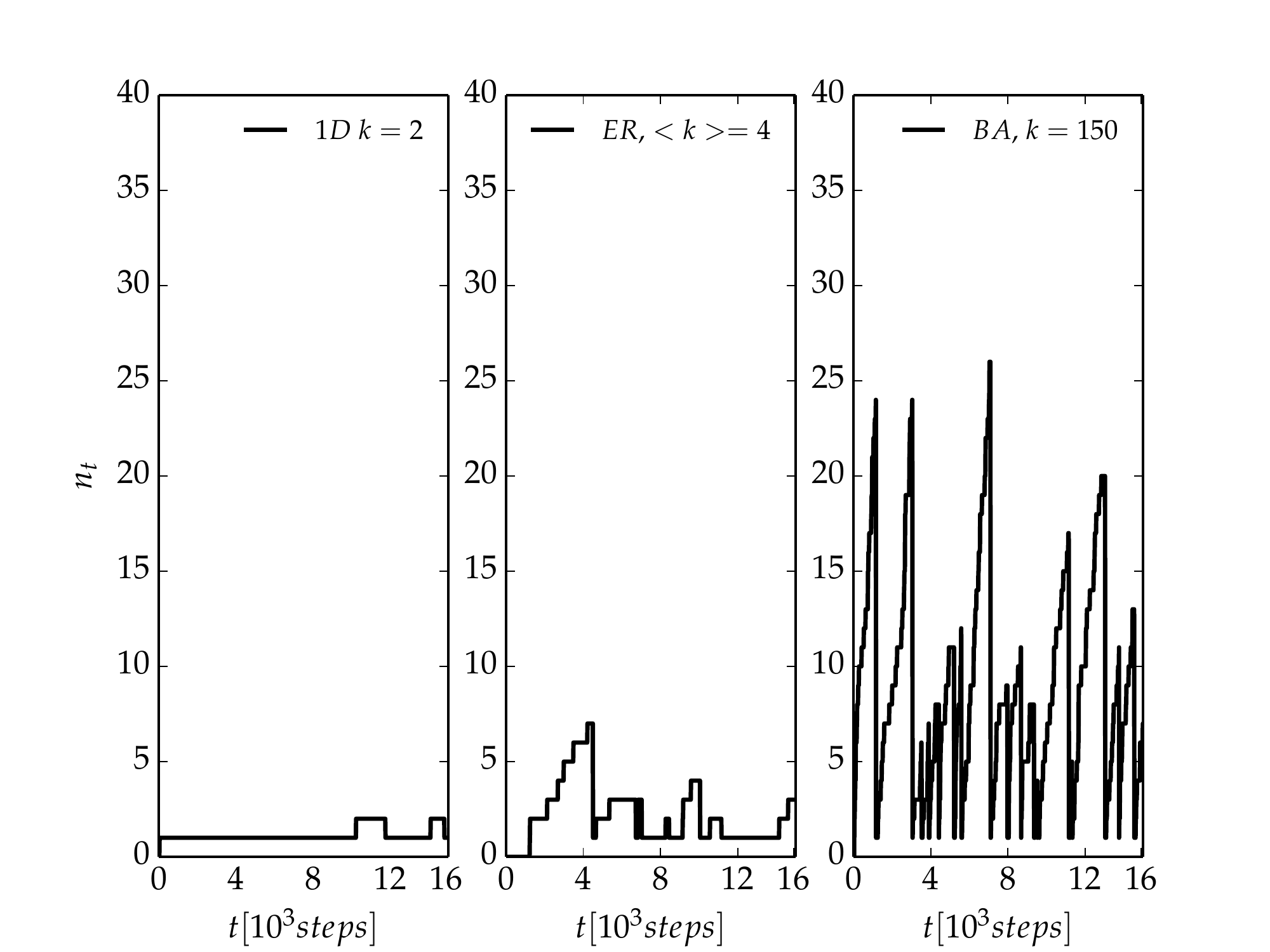}
\caption{Characteristic time-dependence of the inventory's size $n_t$ for some nodes of a 1D lattice (left panel), ER random graph (central panel), and scale-free BA network (right panel), see text for details. 
Both the random graph and the BA networks have average degree $\langle k \rangle =4$. 
The BA network's hub is chosen with degree $k =150$. The curves  $n_t(t)$
are shown for a interval of $1,000$ time-steps.
The model is the (direct) NG and the population size is $N=1,000$ agents (nodes).}
\label{fig:micro}
\end{figure}

While the scaling law for $t_\mathrm{conv}$ proves to be a robust feature for various different complex networks, other dynamical features may be influenced by the topology of the interaction.
For example, investigating the time-evolution of the agents' inventory size, some interesting patterns emerge.
The single agent's microscopic (or internal) dynamics was first studied by Dall'Asta and Baronchelli by means of a master equation method (see Ref.  \cite{DallAsta-2006c}), finding that if $n_t$ is the number of states/words in any given agent's inventory at time $t$, then the form of its distribution  $\mathcal{P}_n\left( k|t\right)$, i.e the probability  that a node of degree $k$ has $n$ states at time $t$, is strongly affected by the network topological properties. 
For instance, far from the consensus, the distribution $\mathcal{P}_n\left( k|t\right)$ takes different forms on homogeneous and heterogeneous networks, due to its dependence on the two first moments $\langle k \rangle$ and $\langle k^{2} \rangle$ of the degree distribution $P(k)$. 
To illustrate this behavior, Fig.~\ref{fig:micro} shows the plot of the inventory size $n_t$ versus time $t$ corresponding to a typical node, in various network architectures, occurring during  the (direct) NG dynamics.

In the left panel of Fig.~\ref{fig:micro}, the inventory size $n_t$ is plotted against time for a given node in a 1D lattice. 
Due to the the coarsening phenomenon, discussed in Sec. \ref{lattices}, the temporal series is indeed bounded, that is, $n_t \leq 2$. 
The behavior of $n_t$ for a typical node of the ER random graph and for a hub of the BA model ($k=150$, $\langle k \rangle =4$) are plotted in the central and right panels of Fig.~\ref{fig:micro}, respectively. 
The curves show that agents/hubs play a much more active role in the semiotic dynamics -- we refer the reader to Ref. \cite{DallAsta-2006c} for further analysis.

In the following, unless explicitly stated otherwise, it is assumed that the direct strategy is adopted (i.e. the direct NG model) and that the NG dynamics takes place on the paradigmatic networks considered above, i.e., networks constructed by means of the Erd\H{o}s-R\'{e}nyi, Barab\'{a}si-Albert,  and Watts-Strogatz network models, as discussed in Sec.~\ref{complexNetworks}.

\begin{figure}
\includegraphics*[scale=0.4]{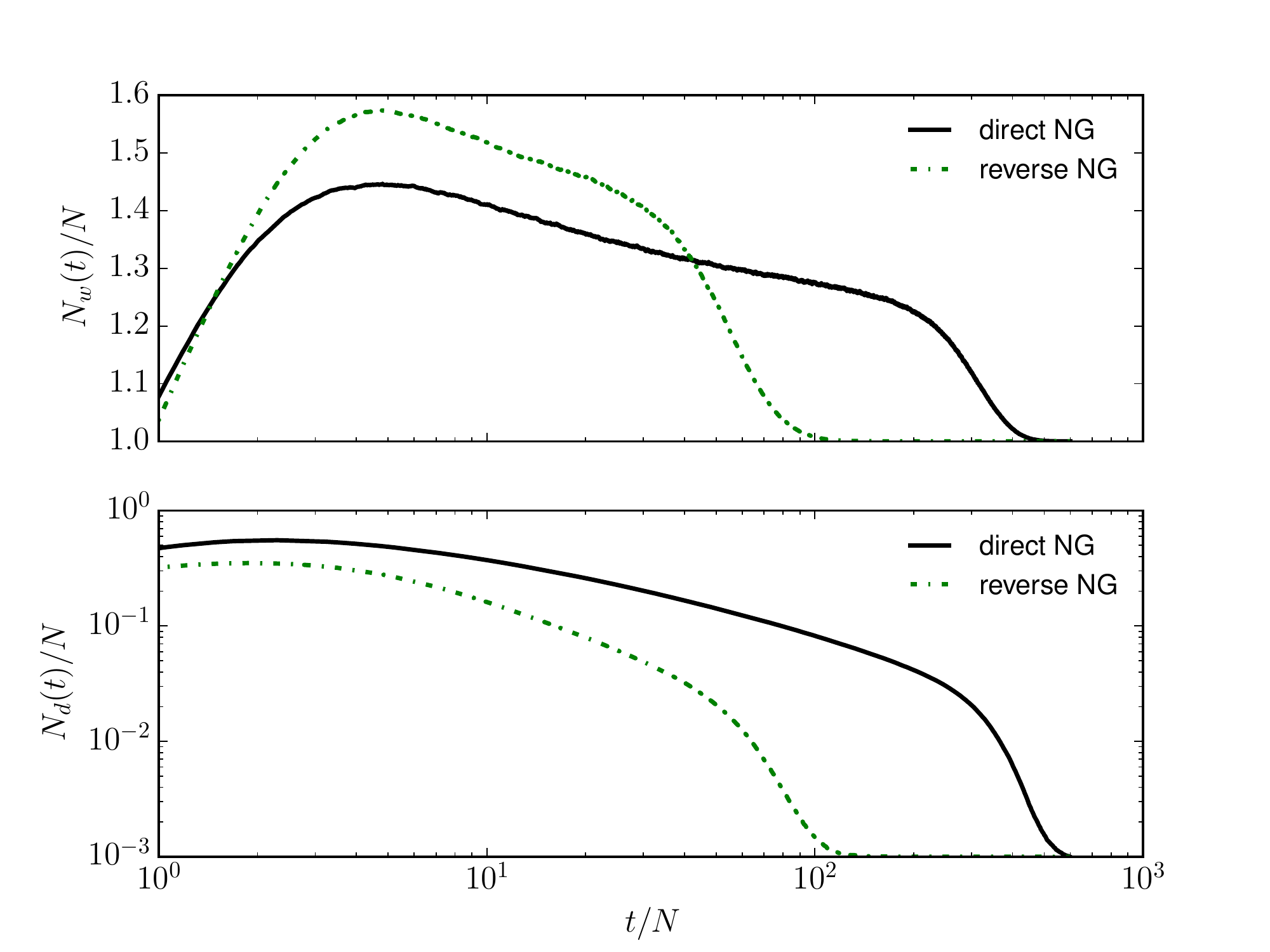}
\caption{Global observables $N_w/N$ (top panel) and $N_d/N$ (bottom panel) as functions of the rescaled time $t/N$, when the direct (solid line) and reverse (dashed line) strategy is adopted.
Here, the population size is $N=1,000$ and the network has a scale-free structure with average degree $\langle k \rangle = 4$ generated through the BA preferential attachment algorithm with parameter $m=4$. 
The curves are obtained by averaging over $600$ realizations. 
See Ref. \cite{DallAsta-2006b} for comparison.}
\label{fig:strategies}
\end{figure}

Finally, we provide an explicit example of dependence on the adopted strategy of agent selection.
In Fig.~\ref{fig:strategies}, the macroscopic observables $N_w$ (top panel) and $N_d$ (bottom panel) on a scale-free network with population size $N=1,000$, average degree $\langle k \rangle = 4$, generated by means of the BA preferential attachment algorithm with $m=4$ (see Sec. \ref{NGonNetworks}), are compared for the cases of a direct and an inverse strategy. 
In the inverse NG model, convergence to consensus is faster, but at the cost of a larger memory usage (larger $N_w$) than in the direct NG model. 
However, the average number of the different words, $N_d(t)/N$, is always smaller in the inverse NG model, due to the fact that hubs, when acting as speakers, convey words to a larger fraction of the population, in turn causing a faster convergence to consensus.
Despite these important differences between the direct and inverse strategies, the scaling law of $t_\mathrm{conv}$ with the system size $N$ remains unchanged\cite{DallAsta-2006b}.

% ===============================================================

\subsection{Basic NG model on small-world networks}
\label{NGonSW}

\begin{figure}
\includegraphics*[scale=0.30]{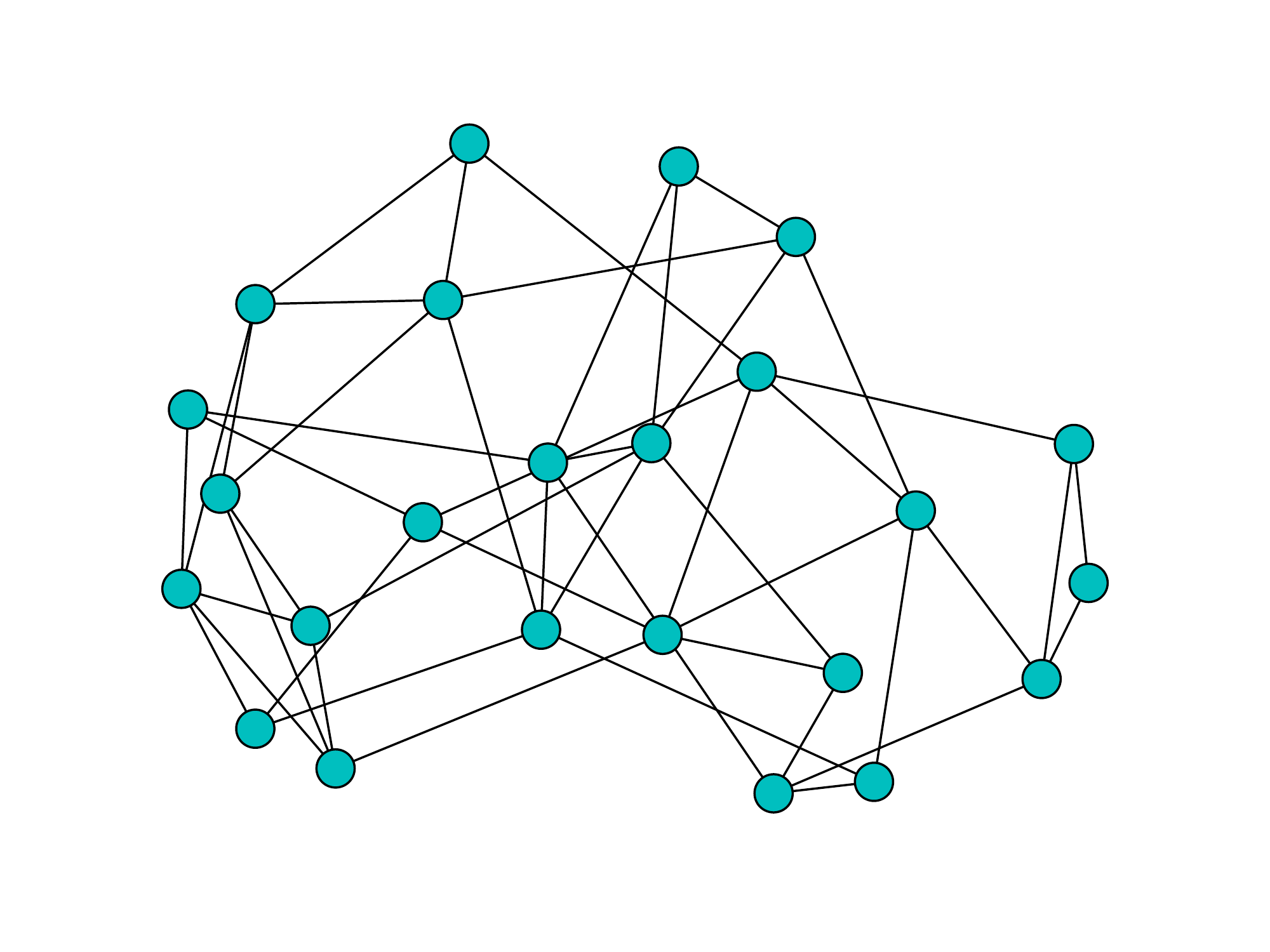}
\caption{Example of network obtained using the WS algorithm for creating a small-world network, starting from a regular ring with $N=24$ nodes connected to four nearest neighbors ($k_i=4$).
Each node was rewired with a probability $p_{\rm WS}=0.4$ (duplicate edges not allowed). 
Note that the network randomness increases with $p_{\rm WS}$. }
\label{fig:wsNetwork}
\end{figure}

\textit{Small-world networks}.
In many biological, social and technological networks, one finds that the connection topology is neither completely regular nor completely random. 
This characteristic feature is observed for instance in the neural network of the worm \emph{Caenorhabditis elegants} and also in the collaboration network of actors in Hollywood \cite{reka2002}. 
Based on this observation, Watts and Strogatz proposed the small-world network model, which can interpolate smoothly between regular and random networks \cite{watts-1998}. 
The Watts and Strogatz (WS) model is called ``small-world'' network because it is high clustered like regular lattices but at the same time shows small path lengths, a typical  feature of random  graphs \cite{watts-1998}. 
In other words, WS graphs can be understood as a superposition of regular lattices and random graphs \cite{Dorogovtsev-2003}. 
The basic algorithm for constructing a WS small-world network is the following\cite{watts-1998, Pastor-Satorras-2015}:
\begin{itemize}
\item[$-$]  Start with a ring lattice, with $N$ nodes and $k_i$ edges per vertex $i$ ($k_i/2$ for each sides).
\item[$-$]  Randomly rewire each edge with a given probability $p_{\rm WS}$, avoiding duplicate edges and self-connections.
\end{itemize}
The required conditions for creating small-world networks of interest are $N \gg  k_i \gg \ln \left(N\right)  \gg 1 $, where the intermediate inequality guarantees that the random graph  will be connected \cite{watts-1998}.
Figure \ref{fig:wsNetwork} shows an example of WS small-world network, that is constructed starting from a regular ring with $N =24$ nodes, each node $i$-th being initially connected to its $k_i=4$ nearest neighbors. 
The corresponding links are then  rewired randomly with probability $p_{\rm WS}=0.4$.

The small-world networks have the following characteristic properties.
First, in the limiting cases of small or large rewiring probabilities, $p_{\rm WS} \rightarrow 0$ or $1$, their clustering coefficient $\langle c \rangle$ and characteristic path length $\langle \ell \rangle$ converge to those of a regular lattice or random graph, respectively. 
However, for a broad interval of probabilities, they form highly clustered networks with a large $\langle c \rangle$, as typical of regular lattices but, at the same time, an  $\langle \ell \rangle$ comparable to those of random graphs \cite{watts-1998}. 
Furthermore, their degree distributions are similar to those of  random graphs, showing an exponential decay for large $k$ values, so that nodes have approximately the same number of edges \cite{reka2002}. 
Therefore, the WS model has the capability to generate complex networks that are relatively homogeneous in their topology\cite{reka2002}.

%%%%%%%%%%%%%%%%%%%%%%%%%%%%%%%%%%%%%%%%%%%%%%%%%%%%%%%%%%%%%%%%%%%%%%
\vspace{0.25cm}

The first numerical investigation of the basic NG dynamics on the small-world  networks was performed by Dall'Asta et. al. \cite{DallAsta-2006a}. 
As discussed above, the WS model introduces long-range connections in an otherwise regular network, thus allowing agents that were originally far from each other to become neighbors. 
Therefore, it is expected that, depending upon the rewiring probability $p_{WS}$, the WS model will introduce a trade-off between the dynamics on a $1D$-lattice and that on a complete graph.

\begin{figure}
\includegraphics*[scale=0.4]{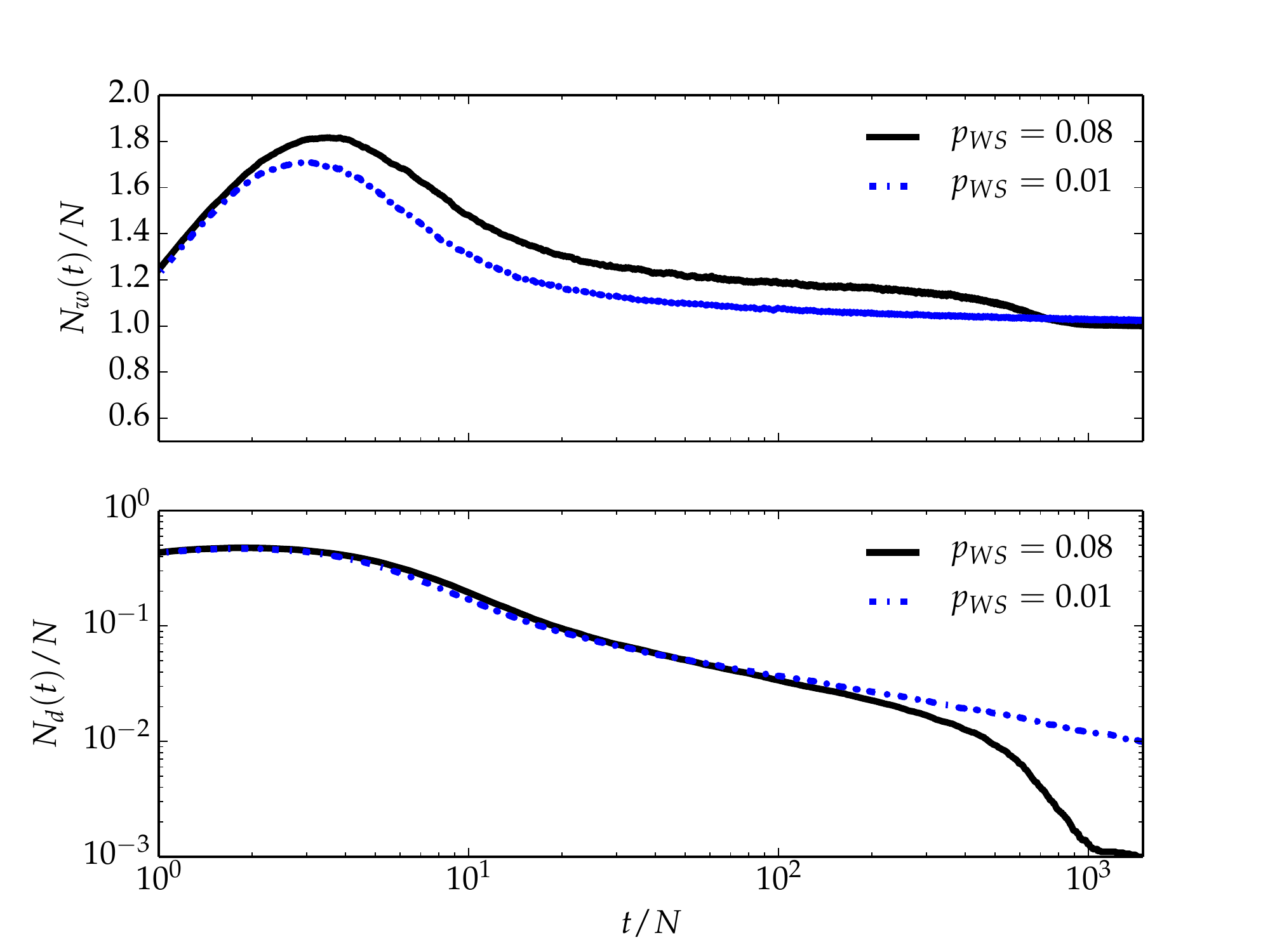}
\caption{Observables $N_w/N$ (top panel) and $N_d/N$ (bottom panel) as functions of the rescaled time $t/N$ for a WS network with average degree $\langle k \rangle = 8$ and for the rewiring probabilities $p_{WS}=0.01$ (dashed line) and $p_{WS}=0.08$ (solid line).  
The curves were obtained averaging over $50$ realizations with a population of $N=1,000$ agents (nodes).}
\label{fig:wsDynamics}
\end{figure}

The effects of a rewiring probability $p_{WS} > 0$ are shown in Fig.~\ref{fig:wsDynamics} for the average memory per agent in the system  $N_w/N$ (top panel) and the average number of words $N_d/N$ (bottom panel) in a WS network model with average degree $\langle k \rangle = 8$, $N=1,000$ nodes, for two different rewiring probabilities, $p_{WS}=0.01$ and $p_{WS}=0.08$. 
These values were chosen according to the condition $1/N \ll p_{WS}  \ll 1$ that guarantees the emergence of a small-world structure through the WS algorithm. 
The curves show that increasing the rewiring probability $p_{WS}$, a faster convergence to consensus is achieved. 
Note that in the time range of Fig.~\ref{fig:wsDynamics}   the approach to consensus is visible only for the case $p_{WS}=0.08$ at $t_\mathrm{conv} \simeq 8 \times 10^{5}$.
Moreover, it is found that increasing the system size $N$, the plateau of the curve of $N_w\left(t \right)/N$  becomes wider\cite{DallAsta-2006a}. 
The dynamics presents a crossover between the typical coarsening phenomenon of 1D-lattice, discussed in  Sec. \ref{lattices}, and the fully-connected regime, see Sec. \ref{ngdynamics}.  
The reason is that the clusters' size grows as $\left(t/N \right)^{1/2}$, while the distance between the short-cuts introduces by the WS model is of order of $ 1/p_{WS}$, so that when they become comparable, at a crossover time $t_{cross} \sim N/p_{WS}^2$, a dynamics similar to that of a fully-connected network emerges. 
This explains why the convergence to consensus in small-world network is much faster than that observed in a 1D-lattice. 
Indeed, it is found that $t_\mathrm{conv} \sim N^{1.4}$, which is a power law similar to that of fully-connected networks, see Table~\ref{table:table1} for comparison, while only a finite memory per agent is required, see Table ~\ref{table:table2}, in contrast to the case of fully connected networks, where it behaves as $\mathcal{O}(N^{1/2})$. 
The latter dynamical feature explains why the curves on the top panel display so long plateaus. 
We refer the reader to Refs. \cite{DallAsta-2006a, Barrat-2007a} for an extensive analysis of the  NG dynamics on small-world networks.
The significant scaling laws for the case of the WS model are reported in Table ~\ref{table:table2}.

%%%  SMALL-WORLD GEOGRAPHICAL NETWORKS %%%%%%%%%%%%%%%%%%%%%%%%%%%

\subsection{Basic NG model on small-world geographical networks}

We briefly recall that the NG dynamics was also studied on the small-world geographical (SWG) networks by Liu et al.\cite{Liu-2009a}.

The SWG networks are constructed by randomly adding links with a fixed geographical distance $d_{ij}$ to $2D$-regular lattices. 
The distance between two nodes is defined as $d_{ij} = |x_i - x_j| + |y_i - y_j|$, 
where $\left(x_i,y_i\right)$ and $\left(x_j,y_j\right)$ denote the Cartesian coordinates of the lattice nodes. 
In their  study, Liu and co-workers investigated how the consensus dynamics, in particular the convergence time of $t_\mathrm{conv}$, depends on the quantity $d_{ij}$, finding a non-monotonic dependence, both for direct and reverse NG strategy.
They also found that the average path (average topological distance) $\langle\ell\rangle$ of the network, which in turn depends on the number of shortcuts introduced, strongly affects $t_\mathrm{conv}$.

% ===============================================================
\subsection{Basic NG model on random geometric graphs}
\label{NGonRandomGeometric}

\textit{Random geometric graphs}.
A typical random geometric graph (RGG) can be generated in the following way \cite{penrose-2003}: 
\begin{enumerate}
    \item Consider a square area of size $L\times L$.
    \item Choose $N$ points uniformly placed on it at random.
    \item Given any two points, if their  distance is less than a given radius (or radio range) $R$, they will be connected.
\end{enumerate}
It is found that a large connected component of the RGGs emerges if the average degree $\langle k \rangle$ becomes greater than a certain critical value $\langle k \rangle_c$.
In the case of a 2D RGG, the critical degree is given by $\langle k \rangle_c \approx 4.5$. 
Moreover, in such a network the connectivity can be tuned, because $\langle k \rangle = \rho \pi R^{2}$, where $\rho = N/L^{2}$ is the density of nodes \cite{Lu-2008a}. 
Note that it is also possible to generate small-world RRGs, starting from a given RRG and simply adding shortcuts between randomly chosen nodes with a given probability. 
In limiting cases, i.e. small rewiring probability and large network size, such small-world RGGs have the same properties of those generated according to the WS algorithm \cite{newman1999-a}.

%%%%%%%%%%%%%%%%%%%%%%%%%%%%%%%%%%%%%%%%%%%%%%%%%%%%%%%%%%
\vspace{0.25cm}

The study of the NG on an RGG can find relevant applications  of technological interest for modeling efficient networks of sensors \cite{Lu-2008a}. 
The typical scenario would be that of a certain number of autonomously operating wireless sensors randomly scattered in large region, whose environment is unknown.
The expected topology of the sensor network similar that of the RGGs.
In such a situation, it is desirable that the system of sensors could develop a common classification or tagging scheme autonomously, without external interventions.

To this aim, Lu et al. \cite{lu-2006a,Lu-2008a} studied a version of NG with local broadcast instead of pairwise communications, since such a version can potentially be a mechanism for leader election among a network of mobile or static sensors placed in a previously unknown environment \cite{Lu-2008a}.
The broadcast process substitutes the usual NG  pairwise interaction with a process where the speaker conveys the selected word to all the neighbors, which form a set of simultaneous multiple hearers. 
The response of these hearers is the usual one of the basic NG: 
if the hearer already had the conveyed word in the dictionary, an agreement process takes place, an event that represents a local success. 
The speaker updates the inventory (i.e shrinks it to the selected word) only if at least one of the hearers had that word  \cite{Lu-2008a}. 
This NG versions on a 2D RGGs presents the same characteristic dynamical features of the coarsening phenomenon described in Sec. \ref{lattices}.
Instead, the convergence time $t_\mathrm{conv}$  is strongly reduced, similarly to what one would expect for the semiotic dynamics on WS model networks (Sec. \ref{NGonSW}).
We refer the reader to Ref. \cite{Lu-2008a} for details. 
Note that within the NG framework, a natural efficient-broadcasting scheme was also proposed by Baronchelli \cite{Baronchelli-2011a} -- we shall return to this version in Sec. \ref{variant1}.

%%%%%%%%%%%%%%%%%%%%%%%%%%%%%%%%%%%%%%%%%%%%%%%%%%%%%%%%%%%%%%%%%%%%%%%%%
\section{Modified NG models}
\label{scenarios}
%\textcolor{red}{Rewritten on 30 March 2020}
%\textcolor{blue}{[OK]}

In this section we mention some additional modified versions of the NG model,
%that have been proposed over the years.  In general, such variants 
which differ from the basic model in the game rules or for the presence of some additional free parameters. 
In general, these NG models are effective for engineering the global consensus. 
We refer the reader to the recent monograph by Chen and Lou \cite{chen-2019a} for an overview of these models.

% ===============================================================
\subsection{NG model restricted to two conventions}
\label{beta_Loreto}

Here and in the following, the term \emph{convention} is used for the NG model in a sense equivalent to that of \emph{name}.

An extension of the NG model was introduced by Baronchelli et al.\cite{Baronchelli-2007b,Loreto_2011a}, in order to take into account the fact that agents may be undecided whether to learn a new name or not.
Depending on the specific application of the NG model, such a feature can well describe also a resilient attitude with respect to cultural changes, a preference of an agent to use the original language, or a random factor that can interfere with the agreement process. 
The generalized model is obtained by introducing a new parameter $\beta$ governing the agreement process and the consequent update of the agents' inventories.
This model is usually referred to as the \emph{2c-NG model} or generalized \emph{$\beta$-model}.
The game rules for a single pairwise interaction between speaker and hearer are re-defined in the following way\cite{Baronchelli-2007b}:
\begin{enumerate}
\item The speaker randomly retrieves a word from its inventory or, if its inventory is empty, invents a new word and adds it to the inventory.
\item. The speaker conveys the selected word to the hearer.
    \begin{itemize}
    \item If the hearer's inventory contains the conveyed word, then
        \begin{itemize}
        \item with probability $\beta$ the basic agreement process takes place (both the agents erase all the words except the conveyed one);
        \item with probability $1 - \beta$, nothing happens.
        \end{itemize}
    \item otherwise, if the conveyed word is missing from the hearer's inventory, the basic one-shot learning process takes place and the hearer adds the new word to the inventory.
    \end{itemize}
\end{enumerate}
Thus, the main change is in the inhibition of the agreement process with a probability $\beta$ .
This model becomes equivalent to the basic NG for $\beta \to 1$.

Notice that since the parameter $\beta$ can be thought as representing the effect of noise in the system, one might expect that, in analogy with  other models, such as the kinetic Ising model, it can be a source of non-equilibrium phase transitions \cite{Odor-2004a}. 
Indeed, it is found that a first order non-equilibrium transition occurs between the absorbing consensus state and an active polarized state, where the population is split in evolving fractions (clusters) with one or multiple names. 
Moreover, their corresponding densities fluctuate around some average values \cite{Baronchelli-2007b, Loreto_2011a}.

\begin{figure}
%\begin{center}
\includegraphics*[scale=1.2]{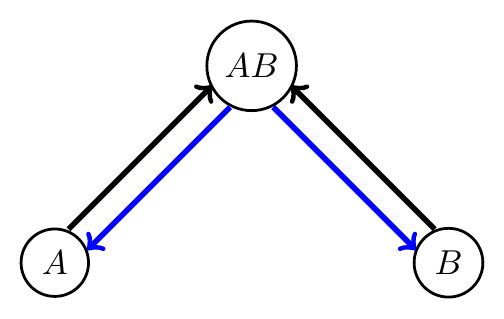}
%\end{center}
\caption{Model scheme with two non-excluding options.
Arrows indicate allowed transitions between the ``bilingual'' state ($A$,$B$) and the ``monolingual'' states $A$ and $B$, see Ref. \cite{Patriarca2012a}.
Direct $A \leftrightarrow B$ transitions are not possible.}
\label{fig:figAB}
\end{figure}

Without loss of generality, this non-equilibrium transition and the associated critical parameter value $\beta_c$, at which it occurs, can be studied considering a model with a finite memory (where agents can have only a finite number of names in their inventories). 
In particular, for the low-dimensional model (the 2c-NG model) with three states $A$, $B$, and $AB$, whose  corresponding transitions are schematically illustrated in Fig. \ref{fig:figAB}, it is straightforward to obtain the evolution equations of the dynamical system in the MF approximation and, from them, the critical value $\beta_c$.
Given the individual transition rates $p_{i \rightarrow j}$, with $i, j = A, B, AB$ (see also Table  \ref{table:table3})
\begin{align}
\label{eq:learningP}
& p_{A \rightarrow AB} =  n_B + \frac{1}{2}  n_{AB} \, ,
%\qquad 
& p_{B \rightarrow AB} =  n_A + \frac{1}{2} n_{AB} \, ,~~~~
\\
\label{eq:agreementP}
& p_{AB \rightarrow A} = \frac{3 \beta }{2}n_A +  \beta  n_{AB} \, ,
%\qquad 
& p_{AB \rightarrow B} = \frac{3 \beta }{2} n_B +   \beta n_{AB} \, ,
\end{align}
where $n_A$, $n_B$, and $n_{AB} \equiv 1 - n_{A} - n_{B}$ are the fractions of population who have only name $A$, only name $B$, and both names $A,B$ in their inventories, respectively, one finds\cite{Baronchelli-2007b}
\begin{align}
\label{eq:firstEq}
&\dot{n}_A = - n_A n_B + \beta n_{AB}^{2} + \frac{3\beta -1}{2} n_A n_{AB}  \, ,
\\
\label{eq:secondEq}
&\dot{n}_B = - n_A n_B + \beta n_{AB}^{2} + \frac{3\beta -1}{2} n_B n_{AB}  \, ,
\end{align}
where the dot represents the time derivative. 
The dynamical system determined by these equations has three equilibrium states (or fixed points): 
$\boldsymbol{n}_1 = (n_A,n_B,n_{AB}) = (1,0,0)$,  
$\boldsymbol{n}_2 = (n_A,n_B,n_{AB}) = (0,1,0)$, and 
$\boldsymbol{n}_3 = (n_A,n_B,n_{AB}) = (n_\beta,n_\beta,1 - 2n_\beta)$, 
where $n_\beta=f(\beta)$ is a function of the parameter $\beta$ (for details see Ref. \cite{Baronchelli-2007b}). 
Moreover, by means of a linear stability analysis, one finds a critical value $\beta_c=1/3$:  
for $\beta > \beta_c $ there is always consensus, either in the state $\boldsymbol{n}_1$ or $\boldsymbol{n}_2$, and the equilibrium state $\boldsymbol{n}_3$ is unstable;
while, for $\beta < \beta_c$, the states $\boldsymbol{n}_1$ and $\boldsymbol{n}_2$ become unstable and the two different words and only  the equilibrium state $\boldsymbol{n}_3$ is stable.

The convergence time diverges as $t_\mathrm{conv} \sim \left(\beta - \beta_c\right)^{-1}$, for $\beta \to \beta_c$ .
Remarkably, it was found that this non-equilibrium phase transition still occurs on a heterogeneous complex network \cite{Baronchelli-2007b}.
We mention that Brigatti and Hern\'andez have studied the discontinuous phase transition of this model on a 2D lattice  \cite{Brigatti-2016a} and, by means of a finite-size scaling analysis, found that the critical value $\beta_c$ is close to $1/3$ when extrapolated in the thermodynamic limit.

\begin{figure}
\includegraphics*[scale=0.45]{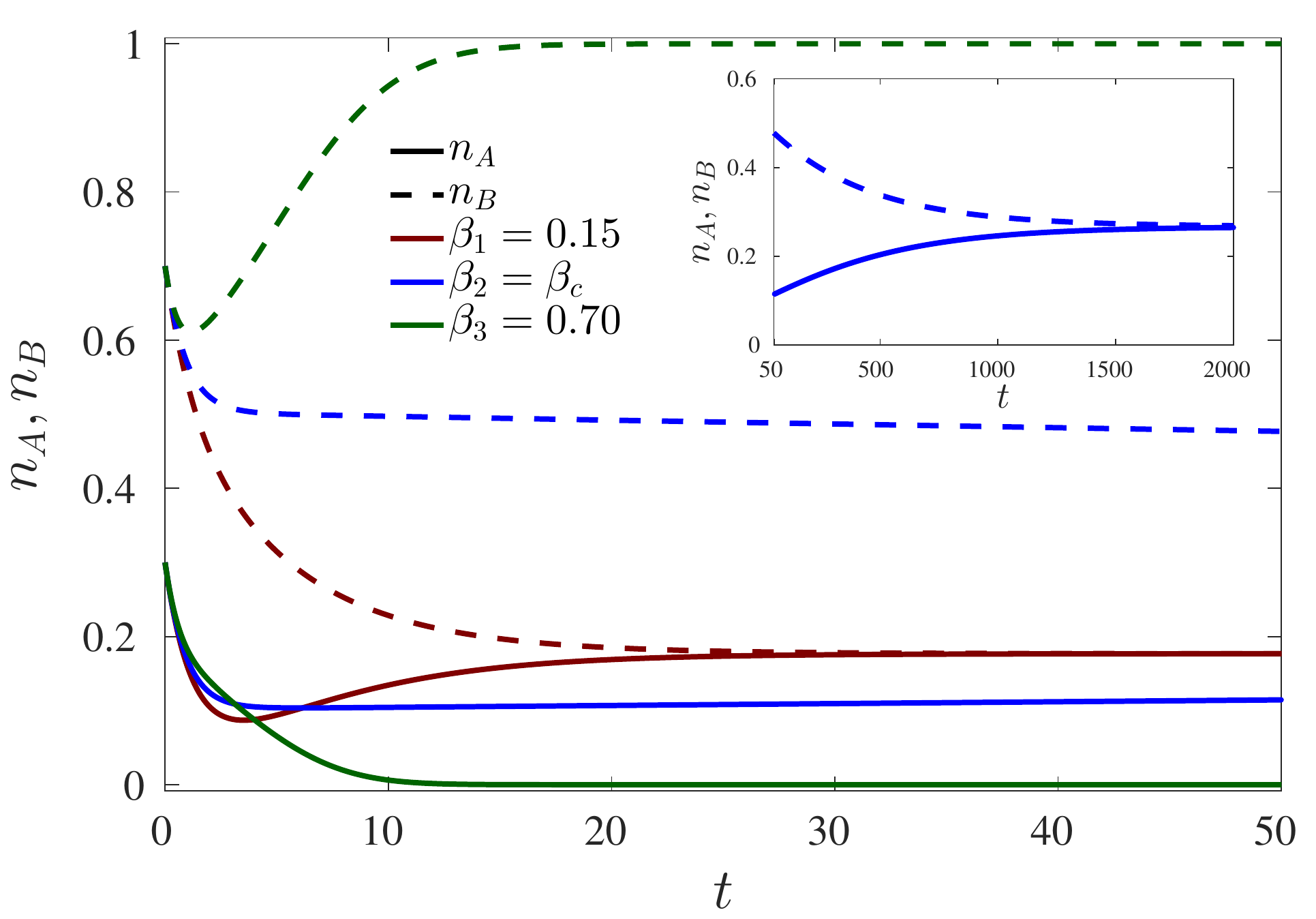}
\caption{Time evolution of $n_A (t), n_B(t)$ (time in arbitrary units) obtained from the numerical solutions of Eqs.~\eqref{eq:firstEq}-\eqref{eq:secondEq} with initial conditions $n_A (0)=0.30$, $n_B(0)=0.70$, for $\beta = 0.15, 0.33, 0.70$ ($\beta_c\approx0.33$).
Convergence to the state of consensus $n_A=1$, $n_B=0$ is achieved for $\beta=0.70>\beta_c$, but not for $\beta=0.15<\beta_c$. 
The inset shows that the numerical solutions at larger times for $\beta_c \approx 0.33$ behaves as expected from the linear stability analysis of Ref. \cite{Baronchelli-2007b}.}
\label{fig:evolution}
\end{figure}

In Fig. \ref{fig:evolution}, we plot the curves for $n_A$ and $n_B$ as a function of time $t$, obtained integrating numerically  Eqs. \eqref{eq:firstEq}-\eqref{eq:secondEq} using the Runge-Kutta method, starting from initial conditions $n_A (0)= 0.3, n_B(0)=0.7$ (and $n_{AB}(0)=0$), for different $\beta = 0.15, 0.33, 0.70$ (compare $\beta_c \approx 0.33$). 
For $\beta = 0.70 > \beta_c$, the system reaches the consensus state $n_A (0) = 0$, $n_B(0)=1$ (and $n_{AB}= 0$) in a very short time.
%This not the case for the other parameter values. 
Instead, at the critical value $\beta = \beta_c \approx 0.33$, the plot of $n_A, n_B$ against time for large times (inset of Fig. \ref{fig:evolution}) suggests that $n_A \approx n_B$ for $t \to \infty$. 
Numerical integration of the MF equations with initial conditions $n_A (0) = n_B(0) = 0.5$ proves that the system converges to the final state  $n_A=  n_B=0.4$. 
This numerical result agrees with those obtained by means of the linear stability theory\cite{Baronchelli-2007b, Castello-2009a}. 
However, if the fluctuations were taken into account, this scenario would be consistent with an unstable situation. 
In fact, a small perturbation would drive the system out of this state, allowing it to reach the consensus at either $A$ or $B$.

%%%%%%%%%%%%%%%%%%%%%%%%%%%%%%%%%%%%%%%%%%%%%%%%%%%%%%%%%%%%%%%%%%%%%%%%%%%%%

In analogy with spin systems, it is customary to define a ``magnetization'' $M = n_{A} - n_{B}$, so that the time-evolution of the system defined by Eqs. \eqref{eq:firstEq}-\eqref{eq:secondEq} can be recast as a single equation for the magnetization \cite{Castello-2009a},
\begin{equation} \label{eq:thirdEq}
\frac{d M}{d t} =  \frac{3\beta -1}{2}  n_{AB} M  \, .
\end{equation}
This NG model is related to the AB-model\cite{Castello-2009a}, where there are two competing languages, denoted by $A$ and $B$, and a third ``bilingual'' state $AB$. 
In that case, the time-evolution for the magnetization, including the acceptance probability $\beta$ into the AB-model, is given by \cite{Castello-2009a}
\begin{equation} \label{eq:fourthEq}
\frac{d M}{d t} =  \frac{1}{2} \beta n_{AB} M \, .
\end{equation}
Note that the two models are equivalent in the MF approximation \cite{Castello-2009a, Castello-2009a}. 
Indeed, the AB model provides the same dynamical systems, i.e. the same Eqs. \eqref{eq:firstEq}-\eqref{eq:secondEq}, once the time is appropriately rescaled. 
However, despite this analogy, no phase-transition occurs in the  AB-model, due to some significant differences at microscopic level -- see Ref. \cite{Castello-2009a} for a detailed discussion.

\begin{table}
\renewcommand{\arraystretch}{2} 
\caption{Possible game outcomes  in 2c-NG model in the MF approximation, see Ref. \cite{Castello-2009a}. The symbol $q$ denotes the branching probability. Note that basic NG dynamics is recovered once  $\beta$ is set to unity.}
\label{table:table3}
\begin{ruledtabular}
\begin{tabular}{c c c c c c  }
%\hline \hline
S & H & S $\rightarrow $ H &  S &  H  &  $P_{outcome}$   \\ [1ex]
\hline
    $A$  & $A$  &  $A$  & $A$ & $A$  & $1.0$   \\
\hline    
    $A$  & $B$  & $A$   & $A$ & $AB$  & $1.0$ \\
\hline   
    $A$  & $AB$  & $A$   & $A$ & $A$  & $\beta$ \\
  $$  & $$  & ~   & $A$ & $AB$  & $1.0 - \beta$ \\
\hline   
  $AB$  & $A$  & $A \left( q=0.5 \right)$   & $A$ & $A$  & $\beta$ \\
  $$  & $$  & $ $   & $AB$ & $A$  & $1.0 -\beta$ \\
 \cline{3-6} 
  $$  & $$  & $B \left(q=0.5 \right) $   & $AB$ & $AB$  & $1.0$ \\
\hline  
 $AB$  & $AB$  & $A \left(q=0.5 \right) $   & $A$ & $A$  & $\beta$ \\
  $$  & $$  & $ $   & $AB$ & $AB$  & $1.0 -\beta$ \\
 \cline{3-6}
    $$ & $$  & $B \left(q=0.5 \right) $ & $B$   & $B$ & $\beta$   \\ 
   $$ & $$  &  & $AB$   & $AB$ & $1.0 - \beta$   \\ [1ex]
%\hline \hline
\end{tabular}
\end{ruledtabular}
\end{table}

% ===============================================================
\subsection{Role of feedback in the NG  model  } \label{variant1}

\begin{figure}[htp]
\includegraphics*[scale=0.55]{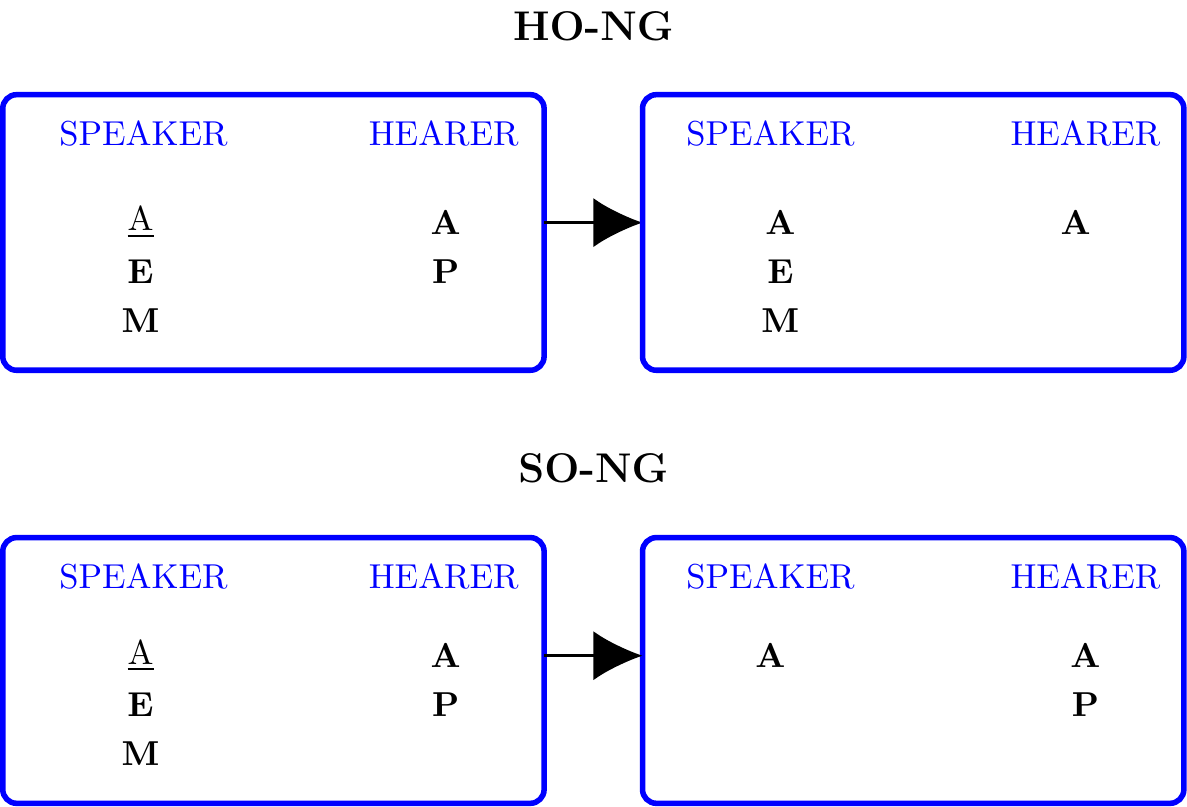}
\caption{Top panel: update of the hearer's inventory after a successful interaction in the modified H0-NG NG model. 
Bottom panel: update of the speaker's inventory after a successful interaction in the modified S0-NG NG model. 
The name conveyed by the speaker is $A$ in both cases.}
\label{fig:asymmetricNG}
\end{figure}

The  NG rules, introduced in Sec.~\ref{ngdynamics}, implicitly require a sort of feedback from the hearer to the speaker in the case of a successful interaction.
In fact, as a consequence of a communication success, \emph{both} the interacting agents update their inventories by reducing them to a single name (the word conveyed) -- no feedback is needed in case of communication failure.
%In this regard, in case of failure no feedback is necessary.
The feedback in the NG model is different from that one can find in the Talking Heads experiments \cite{Steels-1995a} or in Wittgenstein's  linguistic games \cite{Wittgenstein-1953}, which are are real-setting scenarios, where both the agents immediately realize whether or not an interaction is successful. 
These remarks highlight the asymmetric roles played by the speaker and hearer and suggest, as possible modifications of the agreement process of the basic NG, the two following update rules (for the case of success), introduced by Baronchelli\cite{Baronchelli-2011a}:
\begin{itemize}
\item Only the hearer updates its inventory (H0-NG)
\item Only the speaker updates its inventory (S0-NG)
\end{itemize}
These update rules of the agent's inventory after a communication success, in the modified H0-NG and S0-NG models, are illustrated on the top and bottom cartoon in Fig. \ref{fig:asymmetricNG}, respectively.

Baronchelli found that the modified H0-NG NG model gives rise to a fast convergence to consensus, with a scaling law of convergence time with the system size $N$ equivalent to that of the basic NG model (in the MF approximation).
Instead, the convergence time in the modified S0-NG NG model is much longer, see  Table~\ref{table:table4}. 
The reason for the latter peculiar behavior is readily given once one considers the S0-NG model in the context of the generalized $\beta$-model, discussed in the previous Sec. \ref{beta_Loreto}: it can be shown that in that case the evolution of the magnetization $M$ is given by
\begin{equation} \label{eq:so-NG}
\frac{d M}{d t} =  \frac{\beta -1}{2}  n_{AB} M  \, ,
\end{equation}
implying that the corresponding consensus can be reached only if the system is driven by the fluctuations of the magnetization $M$; for the S0-NG dynamics, this occurs in a critical regime\cite{Baronchelli-2011a}, since $\beta_c=1$.

% ----------------------------------------------
\begin{table}
\renewcommand{\arraystretch}{2} 
    \caption{Exponents of the power laws for the  NG dynamics in a fully connected network in H0-NG and S0-NG models  \cite{Baronchelli-2011a}. 
    Here  $t_\mathrm{conv} \sim N^{\alpha}$, $t_\mathrm{max} \sim N^{\beta}$ and $N_{w}^\mathrm{max}\sim N^{\delta}$. }
\label{table:table4}
%\centering
\begin{ruledtabular}
\begin{tabular}{c c c c   }
%\hline \hline
Model & $\alpha$ & $\beta$ &  $\delta$  \\ [1ex]
\hline
    H0-NG     & $1.5 $         & $1.5 $        & $1.5 $   \\

    S0-NG    & $2.0$          & $1.5$         & $1.5$     \\ [1ex]
%\hline \hline
\end{tabular}
\end{ruledtabular}
\end{table}
% ----------------------------------------------

These results suggest that a feedback after a successful interaction between two agents may be not crucial in the context of basic NG models, insofar an efficient convergence to consensus is concerned and if the semiotic model discard the possibility that the homonymy is present\cite{Baronchelli-2011a}. 
In this regard, Baronchelli proposed to employ the modified HO-NG NG model in a broadcasting scheme on complex networks. 
The numerical simulations show that an efficient convergence to consensus can be reached within this scheme on uncorrelated heterogeneous networks generated by uncorrelated configuration model. 
In such a case it is found that $t_\mathrm{conv} \sim N^{\alpha}$, with $\alpha \simeq 1.1$. 
However, the efficiency of this scheme becomes compromised whenever memory consumption is crucial, as memory scales as $N_w^\mathrm{max} \sim N^{\delta}$, with $\delta \simeq 1.1$, which clearly diverges in the thermodynamic limit.

% ===============================================================
\subsection{Other NG models}
\label{variant2}

In the minimal NG model, the size of the agent's inventory is not bounded. 
In principle, any agent could invent and store an unlimited number of different words during the system dynamics. 
However, the invention of new words can occur only at the early stages of the dynamics, due to the increasing overlap of the  inventories. Typically, this would give rise to roughly $N/2$ possible different names, for a system of size $N$, in a fully-connected network. 
In this regard, Brigatti proposed a simple scheme, termed ``open-ended'' NG, which allows agents to continuously invent new words\cite{Brigatti-2008a, Brigatti-2009a, Brigatti-2012a}. 
In this scheme, in case of a communication failure, the speaker is allowed to invent a new word and store it in the inventory. 
The new word is generated according to a Normal Distribution $\mathcal{N}(\mu,\,\sigma)$ where $\mu$ and $\sigma^{2}$ denote mean and standard deviation, respectively.
For sake of simplicity it is usually assumed that $\mu = 0$, while some arbitrary values can be assigned to $\sigma$. 
In general, it is found that this scheme does not hinder the system from reaching  the global consensus.
Interestingly, in the  ``open-ended'' NG, where only the agreement mechanism is present, the system can spontaneously reach an absorbing state  either on fully-connected networks \cite{ Brigatti-2009a} or in 2D lattices  \cite{Crokidakis-2015a}, for some values of the parameter $\sigma$. 
In the latter case, the non-equilibrium phase transition between an absorbing state and a fragmented one occurs at the critical value $\sigma_c \approx 25.6$ \cite{Crokidakis-2015a}. 

Within such an  ``open-ended'' scheme, Brigatti also introduced another NG model, in which each agent's reputation is scored by a time-dependent function $R$. 
The key idea is that in a real-world setting there is always a hierarchical structure between the agents, thereby some agents act as teachers due to their acquired credibility (or reputation). 
In this model, each pairwise interaction updates the quantities $R_S$ and $R_H$, representing the reputation scores of the speaker and hearer, respectively\cite{Brigatti-2008a}.
%, is performed according to the new game rules proposed in Ref. \cite{Brigatti-2008a}. 
Brigatti found that agents with the highest values of $R$ have the capability of spreading their words, which in turn will be found in the final state. 
Moreover, the scaling law for the convergence time with system size presents a novel interesting feature. 
In can be shown that $t_\mathrm{con} = a N^{\alpha_1} + b N^{\alpha_2}$, with $\alpha_1 = 1.2$ and $\alpha_2 = 1.5$  ($a,b$ are some suitable coefficients). 
The functional form of the convergence time, as a linear combination of two different power laws, can be interpreted as a consequence of the presence of two dominant regimes during the dynamics. 
In the first regime, where a faster convergence is expected, there is accumulation of new names in the system, while in the second regime the formation of an hierarchical structure greatly influences the dynamics.
The two scaling regimes are defined by a threshold value $\tilde{N}$ of the system size.
We refer the reader to the original paper for details about the analysis of these results and how the distribution of the values of $R$ assigned to the agents initially affects the macroscopic observables\cite{Brigatti-2008a}.

By appropriately tuning the free model parameters present in the basic NG one can get useful insights on how engineering the consensus in a multi-agent system. In particular, it would be highly desirable to find optimal values of these parameters which could lead to the fastest convergence to the global consensus.
The works of Wang et al. \cite{Wang-2007a},  Yang et al.  \cite{Yang-2008a} and Tang et al. \cite{Tang-2007a}
on the finite-memory NG model, the NG with asymmetric negotiation and connectivity-induced weighted words in  the NG  model respectively come in this perspective.
In the finite-memory NG model, it is assumed that the size of agents' inventories can be a finite tunable model parameter. 
This same model of the NG model, embedded in an ER random graph or a small-word networks, in contrast to the basic NG on the same types of network, relaxes with a time $t_\mathrm{con} $ that has a non-monotonic dependence upon the average degree $\langle k \rangle$. 
Therefore, there exist optimal values of the average degree $\langle k \rangle$ for which a fast convergence to the consensus is achieved.
As long as the other two last modified NG models are concerned, in the model introduced in Ref. \cite{Yang-2008a} the choice of agent $i$, with degree $k_i$, as speaker, is done with a weight $\propto k_i^{\xi}$, depending on a free parameter $\xi$; instead, in the model of Ref. \cite{Tang-2007a}, the choice of the words is done based on the connectivity of the agent $i$, with a weight $\propto k_i^{\xi}$. 
These models crucially depend upon the parameter $\xi$ and, due to the non-monotonic dependence of $t_\mathrm{con}$ upon $\xi$, one can find some optimal values of $\xi$ for which the convergence to consensus on scale-free networks is most efficient.

We conclude this section by recalling some NG models, introduced to investigate semiotics dynamics in the presence of committed agents \cite{Qiming-2009a, Xie-2011a, Xie-2012a, Niu-2017a}. 
Committed agents are individuals whose opinion cannot be changed or, in other words, they are immune to others' influence. 
The modified models produce interesting results, such as a critical size of the committed fraction of agents, corresponds to $\approx 10$\% of the population, beyond which the convergence time $t_\mathrm{con}$ decreases dramatically\cite{Qiming-2009a}.  
Moreover, in the presence of more than a single committed group\cite{ Xie-2012a}, the corresponding phase diagram exhibits new features including bifurcations that can be investigated by means of the theory of dynamical systems \cite{strogatz1995}.

%%%%%%%%%%%%%%%%%%%%%%%%%%%%%%%%%%%%%%%%%%%%%%%%%%%%%%%%%%%%%%%%%%%%%%%%%%%%%%%%

\section{A Bayesian approach to the NG model}\label{sec:bayesianIntro}

In the basic NG model, when the conveyed word is not contained in the hearer's vocabulary, the hearer undergoes a one-shot learning process according to the game rules.
In other words, with a \emph{single} interaction, the new conveyed word is learned by the hearer and added to the hearer's inventory. 
Instead, in real life, the learning process  is typically affected by uncertainties and requires a certain number of positive examples relative to the object concept.%, whose name has to be learnt, over an extended period of time.
The multiple cognitive processes corresponding to such learning experiences eventually allow the learner to generalize (``learn'') the concept. 

In this section, a human learning model is described, which goes beyond the one-shot learning process.
To this aim, a Bayesian framework, which allows a simple and direct coupling between the learning process and the NG model, is used to provide an adequate approach for studying consensus dynamics in a multi-agent system in a real-life setting.

In order to capture the uncertainty of the learning process and take into account the agents' background knowledge, a model based on the Bayesian learning framework developed by Tenenbaum and co-workers\cite{Tenenbaum-1999, Tenenbaum-1999b, Tenenbaum-2000a, Tenenbaum2001, Griffiths2006, Xu-2007a, Tenenbaum-2011a, Perfors-2011a, Lake2015} is appropriate.

\subsection{Bayes' theorem}\label{sec:bayesRule}

Bayesian probability theory provides a rigorous method for inductive inference \cite{Jeffreys1961, Stone2013}, based on Bayes' theorem.
The theorem is named after Thomas Bayes,  an English Presbyterian minister.
However, the actual origin of the theorem is a matter of discussion.
The hypothesis that the theorem was put forward by Thomas Bayes relies on the fact that its formulation was found among Bayes' papers by Richard Price and posthumously published in 1763\cite{bayes-1763}. 
However, the theorem had been stated about a decade earlier in a passage of David Hartley's book \emph{Observations on man} (1749)\cite{Hartley-1749a}, where he writes that an ingenious friend of his communicated the theorem to him. 
More recently, S.M. Stigler managed to garner some evidence\cite{stigler-1993} that Hartley's friend could have been Nicholas Saunderson, a Lucasian professor of mathematics at Cambridge -- but his investigations were not exhaustive enough to exclude Thomas Bayes.

Furthermore, in 1774 Laplace rediscovered and reformulated with more clarity the same theorem.
He applied it to various problems of population statistics, meteorology, geodesy, astronomy (for predicting the mass of Saturn), and even in jurisprudence \cite{Sivia2006}.

Whatever the actual story is, the content of Bayes' theorem can be expressed through the following equation\cite{Stone2013},
\begin{equation}\label{eq:bayesTheorem}
p\left(h|D\right) = \frac{p\left(D|h\right) p\left(h\right)}{p\left(D\right)} \, .
\end{equation}
Here the symbol $h$ represents an hypothesis and $D=\left \{ x_1,  \dots,  x_n \right \}$ a set of observed data.
The theorem provides an expression for the \emph{posterior probability} $p\left(h|D\right)$, that is the conditional probability that hypothesis $h$ is correct given the data set $D$.
The posterior probability is expressed in terms of the product of the conditional probability (or likelihood) $p\left(D|h\right)$ of observing the data set $D$ under the hypothesis $h$ and the \emph{prior probability} $p\left(h\right)$ of the hypothesis $h$.
Finally, the denominator on the right-hand side contains the marginal likelihood (or evidence) $p\left(D\right)$, which can be computed as
\begin{equation}\label{eq:bayesTheorem1}
p\left(D\right) = \sum_{h'} p\left(D|h'\right) p\left(h'\right) \, .
\end{equation}
This quantity is in general difficult to evaluate, but for most of the applications it may be considered as a scaling factor.

Importantly, it is assumed that all the hypotheses $h$ are mutually exclusive and exhaustive in the hypothesis space $\mathcal{H}$. 
Equation \eqref{eq:bayesTheorem} shows that the Bayesian inference is a data-driven process which updates our confidence in a given hypothesis $h$.

Due to the prevailing interpretation of probabilities as frequencies of events (frequentism), the Bayesian inference was discredited for long time \cite{Sivia2006, VanderPlas-2014a, Cox-1946a}, despite having a rigorous mathematical basis \cite{Cox-1946a, deFinetti-1974a}. 
A crucial contribution to rediscovering the Bayes theorem was Sir Harold Jeffreys's book \emph{Theory of Probability} (1939).
The Bayesian inference played an important role during War World  II, for example it was used for cracking the German ciphering machine ``Enigma'' by Alan Turing and co-workers at Bletchley park \cite{Good-1979a}.
By now Bayes' theorem is recognized for its general value and clear logical formulation -- it has been called also ``common sense reduced to calculation'' \cite{Mackay-1999a}.
Even the fictional character of the detective Sherlock Holmes is often credited for reasoning in a Bayesian way\cite{Kadane-2009a, Perfors-2011a}, as in the famous passage\cite{Doyle-2009}
\begin{quotation}
  ‘Once you eliminate the impossible, whatever remains, no matter how improbable, must be the truth’
\end{quotation}

\subsection{The Bayesian learning framework of Tenenbaum}\label{sec:bayesianLearning}

In the following, we discuss the main ideas for casting processes of learning object concepts in a suitable computational Bayesian framework. 
This framework was developed by Joshua Tenenbaum -- we refer the reader to his PhD thesis\cite{Tenenbaum-1999} and some related papers\cite{Tenenbaum2001, Xu-2007a, Tenenbaum-2011a, Perfors-2011a, Murphy-2012a} for an exhaustive discussion.

In his thesis, Tenenbaum ponders the classic problem of induction, observing that humans can generalize informatively from a small number of positive examples. 
This typically human ability can be explained neither by the rule-based approach nor by the similarity-based approach to learning.
Instead, in the case of e.g. machine learning, a large number of training examples is needed -- e.g. Mills et al. recently solved the Schr\"odinger equation for some electrostatic potentials by means of a deep convolutional neural network\cite{Mills-2017a} that requires hundreds thousands training examples for predicting the ground-state energy within chemical accuracy. 
Furthermore, in the case of human learners, negative examples are usually not necessary, while in supervised machine learning the binary classification that roughly corresponds to the same task of concept learning requires both positive and negative examples \cite{Murphy-2012a}.

Let us start by considering how a child learns the meaning of a word, such as  ``cat''.
Typically, somebody around the child will point at a cat (corresponding to a positive examples  ``+'') and utter some sentences -- e.g.  ``Look at the cat!'' or ``This is Andi's cat!''
This kind of situations are reminiscent of the linguistic games discussed by Wittgenstein\cite{Wittgenstein-1953}. 
The process of word learning can be thought equivalent to that of concept learning. 
Here, a concept is understood as a pointer to a subset of entities in the world, also called the concept's extension\cite{Tenenbaum-1999}.

%These considerations and further analyses of the classical problem of induction naturally lead to model concept learning in terms of an inductive inference, where the uncertain character of the learner can be captured within the Bayesian formalism. 
According to Tenenbaum, a computational approach to this problem can be adequately developed from the principles of the Bayesian inference.
To this end, one needs to consider three components of Bayesian inference\cite{Tenenbaum-1999}:
(1) a \emph{likelihood} function, which scores the hypotheses $h$ according the observed examples;
(2) the principle of \emph{hypothesis averaging}; 
and (3) a \emph{prior} distribution over the hypothesis space $\mathcal{H}$.

As for point (1), the strong sampling (generative) model is used, i.e., examples are assumed to be randomly sampled from the true concept $C$.  
In this way, the learner can avoid possible suspicious coincidences arising from the observed examples or data belonging to the data set $D$. 
A direct consequence of this model is that, given the hypothesis $h$ and a set $D=\left \{ x_1,  \dots,  x_n \right \}$ of $n$ examples,  the  corresponding likelihood $p\left(D|h\right)$ reads\cite{Tenenbaum-1999}
\begin{equation}\label{eq:tenenbaum1}
p\left(D|h\right)  = \left(\frac{1}{|h|}\right)^{n} \, .
\end{equation}
Here $|h|$ denotes the size (or measure) of the hypothesis $h$, so that Eq. \eqref{eq:tenenbaum1} encodes the size principle, as it favors the hypotheses with smaller sizes. 
As the preferred hypotheses are the simplest ones, this means that the size principle is equivalent to the Occam's razor \cite{Tenenbaum-1999, Mackay-1999a}.

Point (2) concerns the actual generalization process.
One can notice that by means of the Bayes' theorem, Eq. \eqref{eq:bayesTheorem}, the learner can compute the corresponding posterior probability $p\left(h|D\right)$ for any hypothesis $h$. 
Given a new example $z$, the learner can generalize the concept $C$, i.e. determine whether $z$ belongs to concept $C$'s extension, in accordance with the principle of hypothesis averaging, stating that the generalization function $p\left(z\in C |D\right)$ can be obtained by integrating the predictions over all hypotheses $h$, weighting them with the posterior probabilities $p\left(h|D\right)$\cite{Tenenbaum-1999},
\begin{equation}\label{eq:generalization}
p \left(z \in C | D \right) = \int_{h \in \mathcal{H}}p \left(z \in C | h \right) p\left(h|D\right) dh \, .
\end{equation}
To compute this probability, the previous formalism requires the definition of the hypothesis space $\mathcal{H}$.
Furthermore, a suitable choice of the prior $p\left(h\right)$ for the problem at hand has to be made.
Finally, a natural boundary value $p^*$ determines the condition for the generalization of the concept to take place\cite{Tenenbaum-1999b}:
the agent will generalize only if $p\left(z \in C | D \right) > p^*$.

\begin{figure}
%\begin{center}
\includegraphics*[scale=1.2]{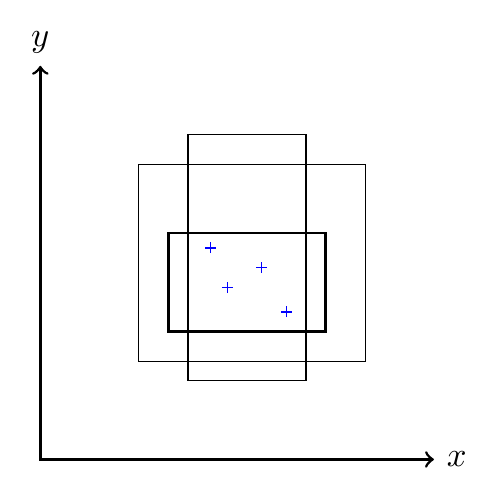}
%\end{center}
\caption{Three different hypotheses, represented as axis-parallel rectangles in the plane $\mathbb{R}^{2}$, and four positive examples ``+'' that are all consistent with the three hypotheses. 
The set of all the axis-parallel rectangles that can be drawn in the plane can be thought as the hypothesis space $\mathcal{H}$.
Figure originally published in Ref. \cite{Marchetti-2020a}. }
\label{fig:hypothesisSpace}
\end{figure}

Without loss of generality, several object concepts can be represented by different geometric patterns\cite{Tenenbaum-2001z}. 
For instance, the concept of ``healthy level'' of an individual in terms of the levels of cholesterol $x$ and insulin $y$ is defined by the ranges $x_a \leq x \leq x_b$ and $y_a \leq y \leq y_b$, where $x_i$ and $y_i$ ($i = a,b$) are suitable values in the Euclidean $x$-$y$ plane $\mathbb{R}^{2}$. 
This means that an axis-parallel rectangle in the plane can be thought as the concept of ``healthy level''. 

In the following, we shall consider a natural hypothesis space $\mathcal{H}$ made up by all the possible axis-parallel rectangles in the plane. 
For instance, Fig.~\ref{fig:hypothesisSpace} shows four positive examples, denoted by the symbol ``+'', associated to four different points of the plane, consistent with three different hypotheses, viewed as axis-parallel rectangles. 
From this cartoon, it is evident how complex the learning process is, due to the presence of many, possibly infinite, axis-parallel rectangles consistent with the same set of examples.

The final step (3), needed for computing the generalization function $p\left(z\in C |D\right)$, is to include the learner's background knowledge through the choice of the prior $p(h)$; in real situations individuals have always some background knowledge.
Such a choice is somehow subjective, an intrinsic feature of the Bayesian inference. 
For the problem of learning an object concept $C$ corresponding to an axis-parallel rectangle, some forms of priors were studied and tested by Tenenbaum \cite{Tenenbaum-1999, Tenenbaum-1999b} in cognitive experiments. 
One of the forms is that of the Erlang (prior) density,
\begin{equation}\label{eq:erlang}
p_E \left(\left( l_1, l_2,  s_1,  s_2 \right) \right) 
    = 
    s_1 s_2  \exp \left \{ -  \left(   \frac{s_1 }{\sigma_1} + \frac{s_2 }{\sigma_2 }      \right)   \right \} \, .
\end{equation}
Here a rectangle of size $s_1 \times s_2$ is represented by the tuple $\left( l_1, l_2,  s_1,  s_2 \right)$, where  $l_1, l_2$ are the Cartesian coordinates of its lower-left corner and $s_1, s_2$ are the sizes along dimension $x$ and $y$, respectively. 
The parameters $\sigma_1, \sigma_2$ in Eq. \eqref{eq:erlang} represent the sizes along dimension $x$ and $y$, respectively, of the true concept rectangle $C$. 
A learner with such a prior does not expect the concept to have sizes much smaller or larger than $\sigma_1$ and $\sigma_2$. 

After inserting Eq. \eqref{eq:erlang} into Eq. \eqref{eq:generalization}, the generalization function $p\left(z \in C | D \right)$ becomes a non-analytical integral that in principle can be computed numerically by Monte Carlo integration\cite{caflisch_1998a}. 
However, for the present case, we use some analytical approximations (upper and lower bounds) obtained by Tenenbaum\cite{Tenenbaum-1999}.
%and they have proven to explain the results obtained from various cognitive experiments, see Ref. \cite{Tenenbaum-1999,Tenenbaum-1999b} for further details.

\subsection{Bayesian naming game model}\label{sec:bayesianNG}

As already mentioned, in a communication of the basic NG model, a one-shot learning process can take place.
The Bayesian learning framework provides a simple way to replace the peculiar one-shot learning process with a realistic model of cognitive process. 
Such a model, here referred to as the \emph{Bayesian naming game} (BNG) model, was introduced in Ref. \cite{Marchetti-2020a}.

The model is restricted to two conventions $A$ and $B$.
As $A, B$ are synonyms in the basic NG model, it is necessary to associate them to a single object concept $C$ in the BNG model. 
Therefore, it is assumed that to the true concept $C$ corresponds the specific axis-parallel rectangle in the Cartesian plane $\mathbb{R}^{2}$ defined by a certain tuple $\left( l_1, l_2,  s_1,  s_2 \right)$, as illustrated above.

In the BNG model, besides the list of the words known, agents are equipped with additional inventories containing the positive examples associated to the corresponding names.
For the model with two conventions $A$ and $B$, each agent $i$ is equipped with three inventories: 
the first inventory (like in the basic NG) is the list $\L_i$ of names known to the agent, which can be $\L_i = [A]$, $\L_i = [B]$, or $\L_i = [A, B]$;
the other inventories, $[+++\dots]_A$ and $[+++\dots]_B$, contain the examples corresponding to the words $A$ and $B$, respectively.

Initially, each agent $i$ is assumed to have either $A$ or $B$ in the list $\L_i$.
If the agent's list contains name $A$ ($B$), then the inventory $[+++\dots]_A$ ($[+++\dots]_B$) contains an initial number of examples  $n_{ex, A}$ ($n_{ex,B}$) associated to $A$ ($B$), while the other corresponding inventory associate to $B$ ($A$) is empty.
The initial examples are points randomly sampled from the rectangle corresponding to the true concept $C$ (strong sampling).

Furthermore, an initial bias in the name learning process is assumed by allowing agents to start generalizing the concept corresponding to $A$ and $B$ only when the sizes $n_{ex,A}(t)$ and $n_{ex,B}(t)$ of the respective inventories reach a threshold number of examples, i.e., when $n_{ex,A}(t) > n^{\ast}_{ex, A}$ and $n_{ex,B}(t) > n^{\ast}_{ex, B}$, respectively. 
If $n^{\ast}_{ex, A} < n^{\ast}_{ex, B}$, then in the early stages of the dynamics it is more likely to learn $A$ than $B$.

\begin{figure}
\includegraphics*[scale=0.5]{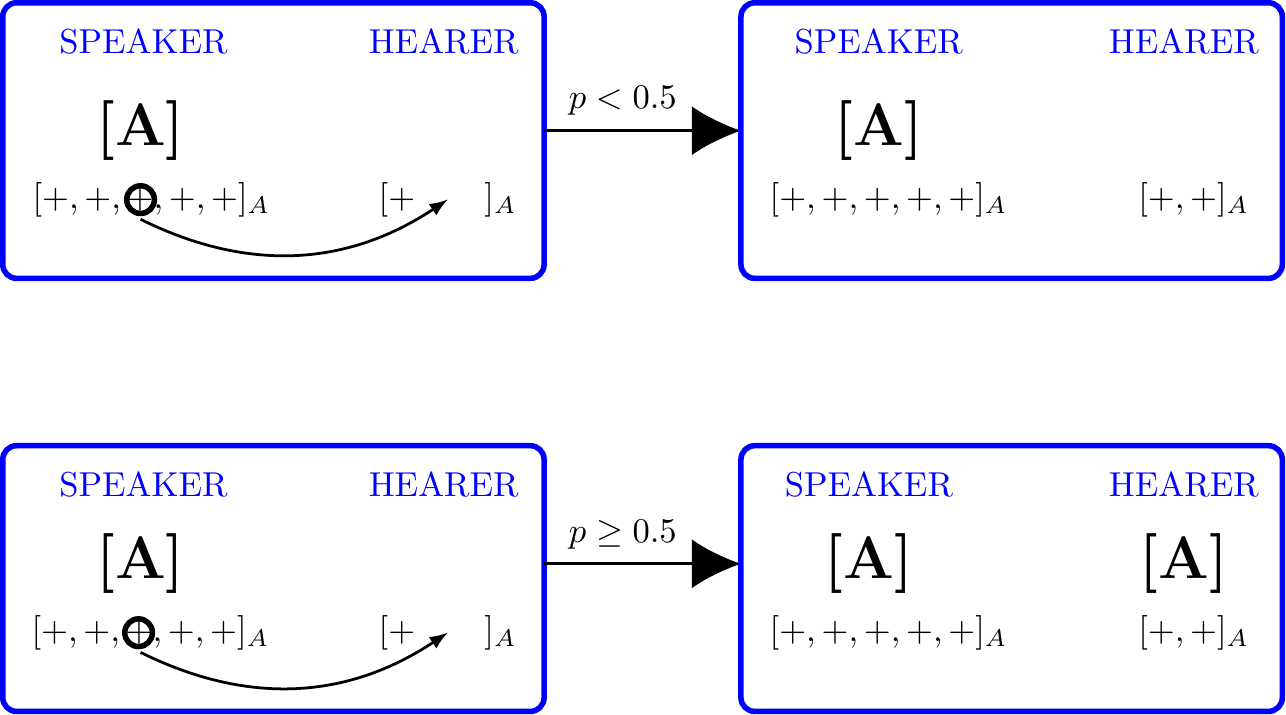}
\caption{Two instances of the new game rules according to the BNG model\cite{Marchetti-2020a}, with critical generalization probability $p^* = 0.5$.
The speaker $i$ conveys name $A$ and shows a corresponding example ``+'' to the hearer $j$, who adds the example to its inventory and then tries to generalize.
Top panel: $p_A < 0.5$, no name is added to the name list $\L_j$. 
Bottom panel: $p_A \ge 0.5$, the hearer successfully generalizes and adds name $A$ to list $\L_j$. 
See text for details.
Figure originally published in Ref. \cite{Marchetti-2020a}.}
\label{fig:bayesianRules}
\end{figure}

The strategy used here to choose the speaker and the hearer is the same of the basic NG model, see Sec. \ref{ngdynamics}.
The game rules for the interacting agents, with agent $i$ in the role of speaker and agent $j$ in the role of hearer, are the following:
\begin{enumerate}
\item The speaker $i$ selects randomly a name from the list $\L_i$ (or the name present if $\L_i$ contains a single name) -- let it be $A$.
\item The speaker $i$ also select randomly an example $z$ among those contained in its corresponding inventory $[+++\dots]_A$. 
\item Then the speaker $i$ conveys the selected example $z$ in association with (e.g. uttering) the selected name $A$ to the hearer $j$.
\item The hearer $j$ adds the new example $z$ (in association with $A$) to its inventory $[+++\dots]_A$. This is a reinforcement process of the hearer's knowledge that always takes place.
\item Depending on the state of the hearer, the following takes place: 
	\begin{enumerate}
		\item \emph{Generalization}. 
		If the selected name $A$ is \emph{not} present in the hearer's list $\L_j$, then the hearer $j$ tries to generalize.
		The outcome depends on the value of $p_A \equiv p(z \in C | X_A)$, given by Eq. \eqref{eq:generalization}, 
		where $X_A = [+++\dots]_A$ is the set of $A$-examples of agent $j$.
		
		    %\item -- \emph{Successful generalization}. 
		    If $p_A \ge p^*$, the hearer successfully generalizes the concept $C$ and connects the inventory $[+++\dots]_A$ to name $A$; also, agent $j$ adds name $A$ to the list $\L_j$.
		    Starting from this moment, agent $j$ can communicate concept $C$ to other agents by conveying an example taken from the inventory $[+++\dots]_A$ while uttering  name $A$.\\
		    %\item 
		    If $p_A < p^*$, the generalization is unsuccessful. 
		
		\item  \emph{Agreement}. If the name $A$, uttered by the speaker $i$, is present in the hearer's list $\L_j$, then an agreement takes place, as in the basic NG model, i.e. both agents $i$ and $j$ keep only $A$ in their name lists $\L_i$ and $\L_j$, removing $B$, if present. 
		However, no examples contained in the inventories of agents $i$ and $j$ are removed.
		\end{enumerate}
\end{enumerate}
Two possible pairwise interactions in the BNG model leading to successful or unsuccessful generalization are illustrated in Fig. \ref{fig:bayesianRules}.

As the Bayesian agents are expected to generalize concept $C$ from a small number of examples, the BNG model should be asymptotically equivalent to the 2c-NG model, however, with a longer average convergence time $t_\mathrm{conv}$.
Also the absorbing states are expected to be the same, i.e. with either convention $A$  ($n_A = 1, n_B = 0, n_{AB}= 0$) or $B$  ($n_A = 0, n_B = 1, n_{AB}= 0$). 
This is confirmed by the linear stability analysis of the corresponding MF equations for the BNG model, which read\cite{Marchetti-2020a} 
\begin{align} 
\label{eq:firstEq1}
&\dot{n}_A = - p_B n_A n_B +  n_{AB}^{2} + \frac{3 - p_B}{2} n_A n_{AB}  \, ,
\\
\label{eq:secondEq1}
&\dot{n}_B = - p_A n_A n_B +  n_{AB}^{2} + \frac{3 - p_A}{2} n_B n_{AB}  \, .
\end{align}
Here $p_A$ and $p_B$ are time-dependent functions, whose numerical values are the outputs of the monotonically increasing time-dependent generalization function $p(t)$\cite{Marchetti-2020a}.
The function $p(t)$ should be understood as an average of Eq. \eqref{eq:generalization} over many possible dynamical realizations. 
Note that apart from the early stages of the dynamics, when $p_A(t) \neq p_B(t)$, due to the initial bias, in general one has that 
$p_A(t) \simeq p_B(t)\simeq  p(t)$, see Ref. \cite{Marchetti-2020a} for details.

%%%%%%%  DATA  %%%%%%%%%%%%%%%%%%%%%%%%%%%%%%%%%%%%%%%%%%%%%%%%%%%%%%%%%%
In the numerical simulations in Ref. \cite{Marchetti-2020a}, 
it was assumed that the initial numbers of examples provided to the agents are $n_{ex, A}=n_{ex, B}=4$;
the critical generalization threshold $p^*= 0.5$;
and the minimum numbers of examples required for generalizing $n^{\ast}_{ex, A} = 5$ and $n^{\ast}_{ex, B} = 6$.
Finally, the tuple defining concept $C$ was assumed to be $\left( 0, 0,  \sigma_1= 3,  \sigma_2= 1 \right)$; 
note that this particular choice does not affect the semiotic dynamics. 
%%%%%%%%%%%%%%%%%%%%%%%%%%%%%%%%%%%%%%%%%%%%%%%%%%%%%%%%%%%%%%%%%

\begin{figure}
%\begin{center}
\includegraphics*[scale=0.45]{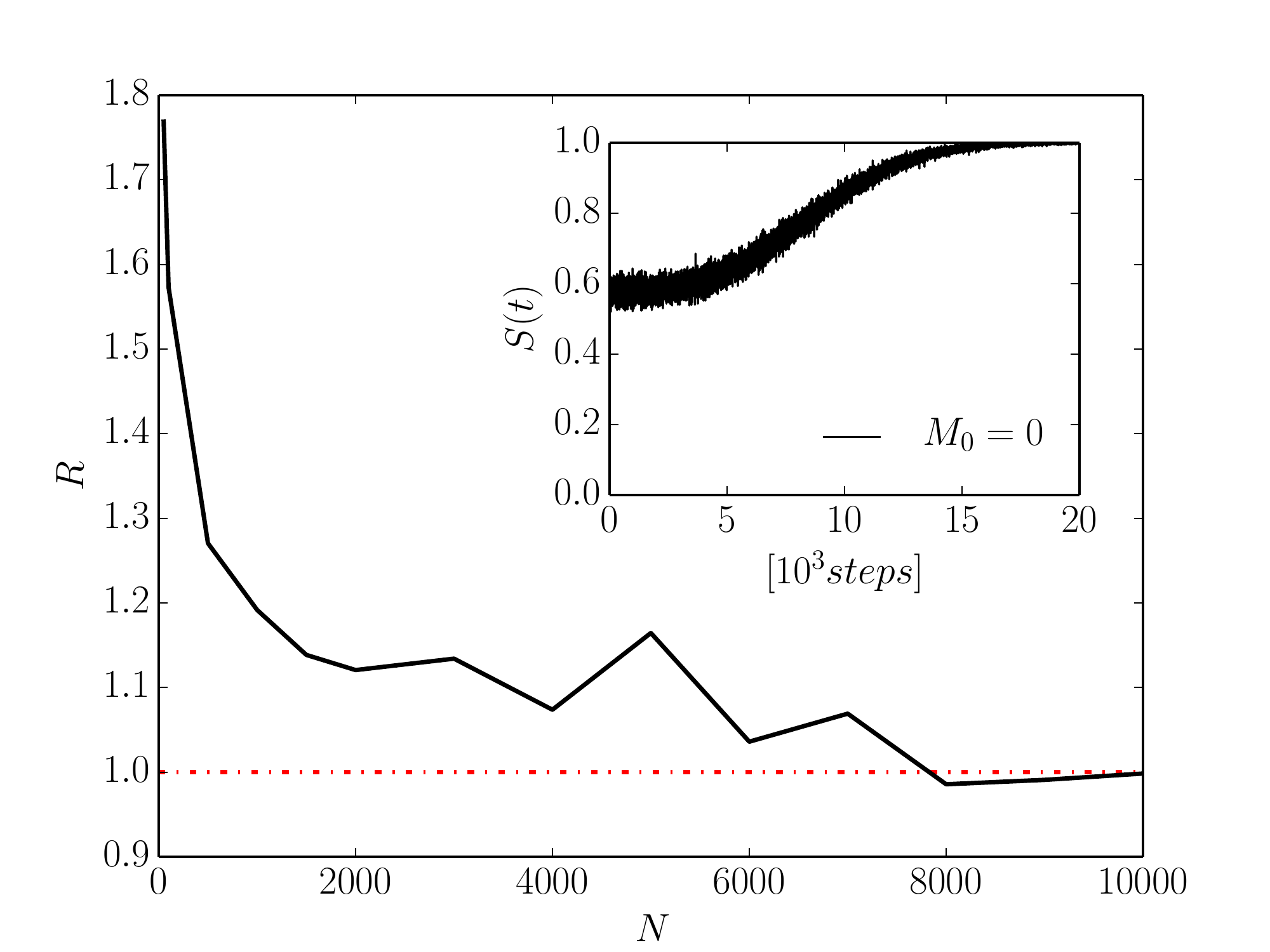}
%\end{center}
  \caption{Ratio $R = t_\mathrm{conv}/\tilde{t}_\mathrm{conv}$,  where $t_\mathrm{conv}$ and $\tilde{t}_\mathrm{conv}$ are the (average) convergence times for the BNG and 2c-NG models, respectively, against the system size $N$, with initial condition $M_0=0$ and averaged over $900$ runs.
  The inset shows the time evolution of the success rate observable  $S(t)$ for $N=1,000$ and initial $M_0=0$ averaged over $600$ runs.
  Figure originally published in Ref. \cite{Marchetti-2020a}.
  }
\label{fig:ratio}
\end{figure}

In Fig. \ref{fig:ratio},  the ratio $R = t_\mathrm{conv}/\tilde{t}_\mathrm{conv}$,  where $t_\mathrm{conv}$ and $\tilde{t}_\mathrm{conv}$ are the (average) convergence times for the BNG and 2c-NG models, respectively, is plotted against the system size $N$, assuming an unpolarized initial state with $M_0=0$.
The ratio $R$ converges to unity for relatively large  numbers of agents, showing that in that limit the two models have the same average convergence times -- in other words the Bayesian learning process becomes equivalent to a one-shoot learning in large systems. 
However, the two models do not have the same dynamics.
The inset of Fig. \ref{fig:ratio} shows the success rate $S(t)$ typically observed in the BNG model against time,  averaged over $900$ realizations of a system of $N=1,000$ agents, starting from an initial state with $M_0=0$. 
This macroscopic observable shows the typical S-shaped curve found in the basic NG model, despite the fact that in the BNG it is assumed that a failure (i.e. $S=0$) occurs only when a hearer fails to  generalize, while $S=1$ both in the case of a reinforcement and of a learning process.

\begin{figure}
\includegraphics*[scale=0.45]{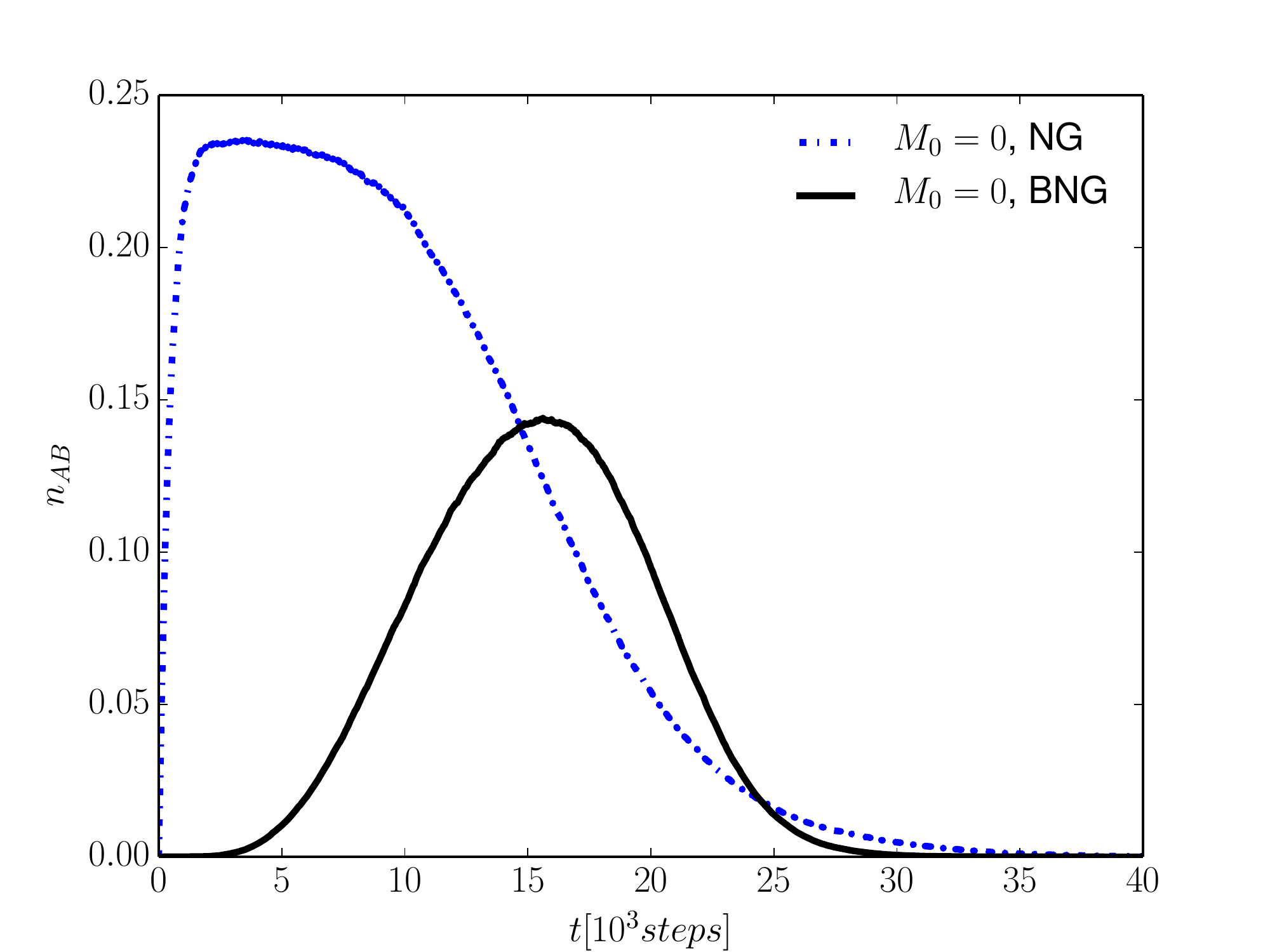}
    \caption{Time evolution of the population fraction $n_{AB}$ for a system with size $N= 1,000$ starting from an unpolarized initial state $M_0=0$ for the NG dynamics (solid line) and the BNG dynamics (dashed line).
    The corresponding curves are obtained by averaging over $600$ runs. 
    The corresponding (average) converge times for the NG and BNG models are $\tilde{t}_\mathrm{conv} \approx 24 \times 10^{3} $ and $t_\mathrm{conv} \approx 29 \times 10^{3}$, respectively.
    Figure originally published in Ref. \cite{Marchetti-2020a}.
    }
\label{fig:nAB}
\end{figure}

From a simple comparison of Eqs. \eqref{eq:firstEq1}-\eqref{eq:secondEq1} with Eqs. \eqref{eq:firstEq}-\eqref{eq:secondEq}, assuming $\beta = 1$, one expects a richer as well as different dynamics in the Bayesian model, with respect to the basic NG model, due to the presence of a new additional and non-trivial temporal component $p(t)$ in Eqs. \eqref{eq:firstEq1}-\eqref{eq:secondEq1}. 
This is indeed the case, as confirmed by Fig. \ref{fig:nAB}, which compares the time evolution of the bilingual fraction $n_{AB}$ of the agents having both names in their name list, in the basic NG and BNG models, starting with an initial state $M_0=0$. 
In the early stage of the dynamics, the name learning  (in real-setting) process prevents agents to add new names to their inventories. 
On the contrary, in the basic NG model, an agent can quickly acquire a new name within a relatively short time (each acquisition is done through a one-shot learning process) allowing the  $n_{AB}$ curve to reach its maximum much earlier.
The following plateau for the case of the basic NG model is not observed in the corresponding $n_{AB}$ curve of the BNG model. 
Moreover, the latter curve exhibits a characteristic bell-shape due to the interplay of the agents' cognitive efforts and of the agreement processes. 
These considerations should be sufficient to demonstrate that these two models give rise to different characteristic non-equilibrium dynamics.

The cognitive feature of the BNG model clearly emerges from the dependence upon the average numbers of the positive examples $\bar{n}_{ex, A}, \bar{n}_{ex, B}$, relative to $A,B$, respectively.
By observing that the average number of pairwise interactions performed by agents until the system reaches the consensus at $t_\mathrm{conv}$ is  $\bar{n}_{int}  = \bar{n}_{ex, A} + \bar{n}_{ex, B} $, one should expect that $t_\mathrm{conv}  \approx  \bar{n}_{int} N $. 
This is confirmed by fitting the numerical results of $t_\mathrm{conv}$ against the system size $N$, averaging over many realizations.  
Table \ref{table:tableBNG} lists the results for a population of $N= 1,000$ agents\cite{Marchetti-2020a}.
The last column lists the possible outcomes at consensus: for initial condition $M_0=0$, both names $A, B$ are likely to be found at consensus, but in general $A$ has more chances and, for a system size $N$ greater than a threshold value, $N > N^{\ast}  \approx 500$, $A$ will always win \cite{Marchetti-2020a}. 
It is also found that $\bar{n}_{int}$ becomes weakly dependent upon the system size $N$ when increasing $N$. 
This dynamical behavior is a direct consequence of the size principle.

\begin{table}
\renewcommand{\arraystretch}{2} 
\caption{Scaling laws for the convergence time, $t_\mathrm{conv} \sim N^{\alpha}$ with the system size $N$.\cite{Marchetti-2020a}
Here the parameters are  $n^{\ast}_{ex,A}=5$,  $n^{\ast}_{ex,B}= 6$, and we consider different initial conditions $M_0=0, M_0=-0.4, M_0=0.4 $. 
The average number of examples, $\bar{n}_{ex, A}, \bar{n}_{ex, B}$, stored at  $t_\mathrm{conv}$, are obtained by averaging over $600$ realizations for a system with $N= 1,000$ agents. 
The last column of the Table shows the possible outcomes at consensus. 
\label{table:tableBNG}}
%\centering
\begin{ruledtabular}
\begin{tabular}{c c c c  c}
%\hline \hline
$M_0$ & $\alpha$  & $\bar{n}_{ex, A}$  &  $\bar{n}_{ex, B}$ & outcome \\ [1ex]
\hline
$0$      & $1.06$          & $ 20$       & $8$   & $A, B$   \\
$-0.4$       & $ 1.08$         & $3$         & $19$  & $B$   \\ 
$0.4$          & $1.09$          & $18$        & $3$   & $A$   \\ [1ex]
%\hline \hline
\end{tabular}
\end{ruledtabular}
\end{table}

\begin{figure}
%\begin{center}
\includegraphics*[scale=0.45]{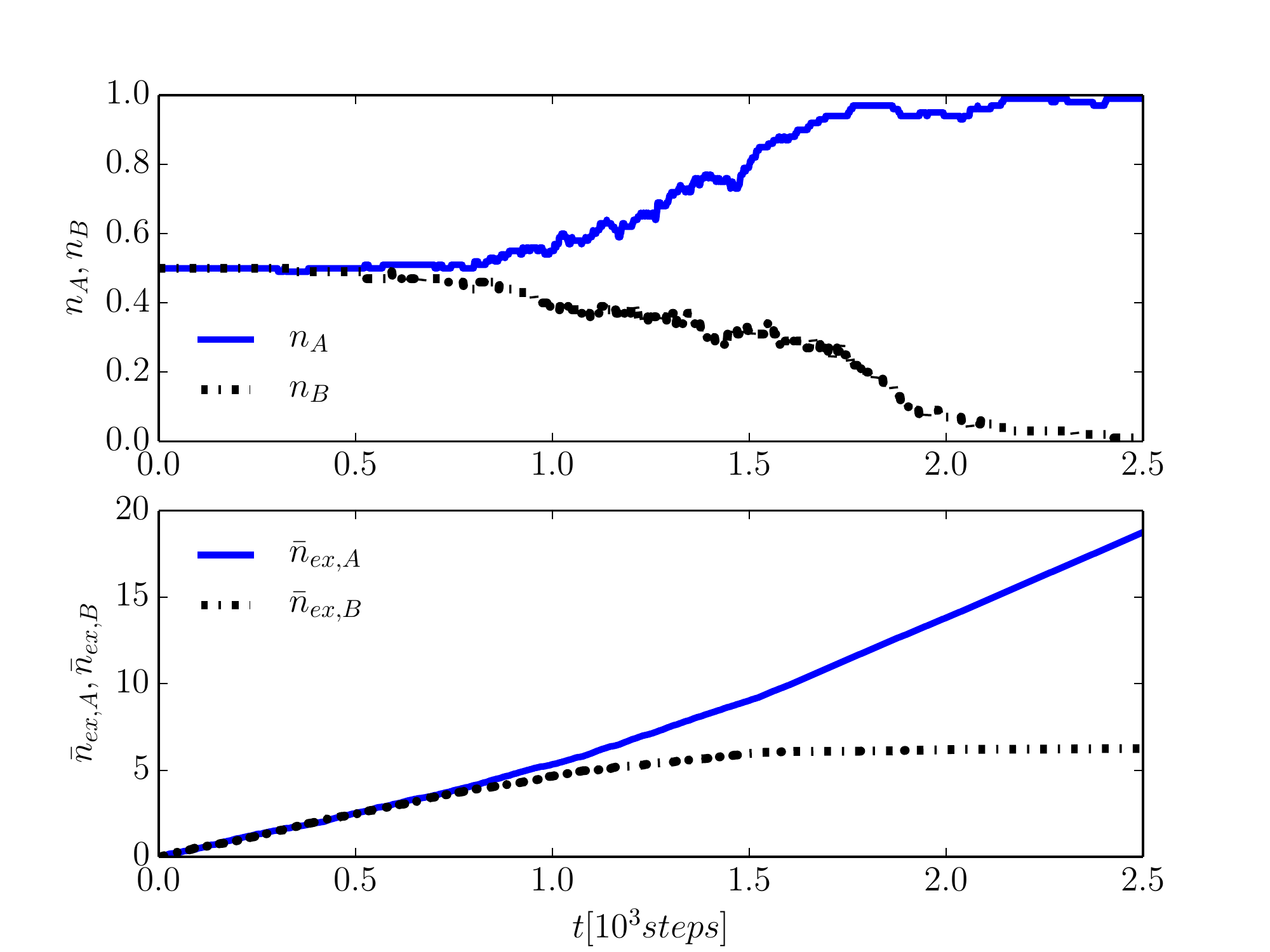}
%\end{center}
    \caption{Top panel: time evolution of the population fractions $n_A, n_B$  for a system of size $N= 100$, obtained from a single realization with initial magnetization $M_0= 0$. Consensus is reached about the convention $A$. 
    Bottom panel: time evolution of the corresponding average numbers of positive  examples $\bar{n}_{ex,A}(t)$ and $\bar{n}_{ex,B}(t)$ recorded by the agents during the dynamics, causing the $n_A$ and the $n_B$ curves to split at the critical time  $t^{\ast} \approx 0.9 \times 10^{3}$.
    Figure originally published in Ref. \cite{Marchetti-2020a}.
}
\label{fig:bifurcation}
\end{figure}

The role of the (average) number of examples can be further investigated considering the time-dependence of the cognitive efforts performed by the agents during the semiotic dynamics.
As these efforts are absent in the basic NG model there is no way to explain why  $A$ or $B$ will be found at consensus. The possible outcomes are always random for the dynamics is purely stochastic. On the contrary, within a Bayesian learning framework, despite the dynamics still remains stochastic, one can guess the outcome at consensus monitoring the agents' cognitive efforts through the time-dependence of $\bar{n}_{ex, A}(t), \bar{n}_{ex, B}(t)$. The reason is that 
the  probabilities $p_A, p_B$ crucially depend on $\bar{n}_{ex, A}(t), \bar{n}_{ex, B}(t)$. In particular,  their instantaneous difference $\delta p(t) \equiv p_A(t) - p_B(t)$  at a certain critical time $t^{\ast}$ would determine which name would be found at consensus.
This is clearly illustrated in Fig. \ref{fig:bifurcation} where the top panel shows the population fractions $n_A, n_B$ against time for a system of size $N= 100$, obtained from a single realization assuming $M_0= 0$. It is shown that after an early stage where $n_A(t) \approx n_B(t)$, there is a critical time $t^{\ast} \approx 0.9 \times 10^{3}$ at which the curves start to differentiate, allowing the whole population to reach the consensus at $A$. In order to explain this dynamical feature, one needs to look at temporal behavior of the average numbers $\bar{n}_{ex,A}(t)$ and $\bar{n}_{ex,B}(t)$ of positive  examples, stored by the agents during the dynamics. 
To this end on the bottom panel of  Fig. \ref{fig:bifurcation}, the quantities $\bar{n}_{ex,A}(t)$ and $\bar{n}_{ex,B}(t)$ are plotted against time. 
In such a case, the initial cognitive bias, i.e.  $n^{\ast}_{ex, A}=5, n^{\ast}_{ex, B} = 6$ and the initial condition $M_0= 0$ favor the learning process of $A$ for  $p_A(t) \gtrsim p_B(t)$ for $t  \gtrsim  t^{\ast}$ as    $\bar{n}_{ex,A}(t)  \gtrsim \bar{n}_{ex,B}(t)$.    
This is again consistent with  the size principle of the Bayesian learning framework, as the probabilities become exponentially greater with the number of examples, see Eq. \eqref{eq:tenenbaum1}. 
These peculiar aspects and other novel features of of the BNG dynamics are thoroughly discussed through a geometrical approach\cite{Arnold-1988a} in Ref. \cite{Marchetti-2020a}. 

 %%%%%%%%%%%%%%%%%%%%%%%%%%%%%%%%%%%%%%%%%%%%%%%%%%%%%%%%%%%%%%%%%%%%%%%
 
 \section{Conclusion}
 \label{conclusions}
 
In this short review, we summarized some NG models, focusing on different underlying topologies, various dynamical rules of the interactions between agents, and different levels of description of the learning process.
The range of examples discussed shows how many different scenarios can be tackled at a quantitative level through suitably modified versions of the NG model.

The dynamics of the basic NG model relies on a simple one-shot learning process (in which new words are learned) and an agreement mechanism (in which unused words are discarded).
The interplay between these two processes allows a simple statistical description of a learning process and generates a dynamical mechanism leading to consensus in a group of interacting individuals.

When more difficult questions are considered, in particular how the generalization of an object concept takes place in the mind of an individual, starting from a set of recorded data, then new frameworks are required.
This problem can be approached noting that, at a phenomenological level, human learning proceeds following some principles of Bayesian inference\cite{Tenenbaum-1999, Tenenbaum-1999b}.
For this reason, we also presented a Bayesian version of the NG model of word learning\cite{Marchetti-2020a}, which is a model of semiotic dynamics that merges the agreement process of the basic NG model with the Bayesian learning framework put forward by Tenenbaum and co-workers\cite{Tenenbaum-1999,Tenenbaum-1999, Tenenbaum-1999b}.

Semiotic dynamics in general and in particular the Bayesian model discussed in this review represent a step forward in the modeling of the human learning process and the corresponding consensus dynamics, realized only through the interactions among different individuals.
It is an interesting feature of the model, characteristic of many complex systems, that the individual learning dynamics and the collective consensus dynamics cannot be disentangled from each other, but it is the interplay of their combined evolutions that shapes the individual concepts into a commonly shared set of notions.

Future research is to be expected to move toward a cognitive dimension of complex systems modeling, for its insight into theoretical questions of language dynamics\cite{patriarca2020}, such as the human word-learning process and the origin of language, and the possible technological applications to the design of systems of intelligent interacting units that have the ability to perform complex functions without external controls.

\section*{Acknowledgments}

The authors acknowledge support from the Estonian Ministry of Education and Research through Institutional Research Funding IUT (IUT39-1), the Estonian Research Council through Grant PUT (PUT1356), and the ERDF (European Development Research Fund) CoE (Center of Excellence) program through Grant TK133.\\
The authors also thank David Navidad Maeso for numerically integrating Eqs. \eqref{eq:firstEq}-\eqref{eq:secondEq}.

%%%%%%%%%%%%%%%%%%%%%%%%%%%%%%%%%%%%%%%%%%%%%%%%%%%%%%%%%%%%%%%%%%%%%%%%%%%%%%%%%%%%%%%%%%%%%%%%%
%\bibliography{LIBRARY_NG-5}
%\bibliographystyle{unsrt} % comment this to have the option "longbibliography* working.
%%%%%%%%%%%%%%%%%%%%%%%%%%%%%%%%%%%%%%%%%%%%%%%%%%%%%%%%%%%%%%%%%%%%%%%%%%%%%%%%%%%%%%%%%%%%%%%%%


\end{document}